\documentclass[12pt]{article}
\usepackage{amsmath,amssymb,amsfonts}
\usepackage[dvips]{graphicx}
\usepackage{subfigure}
\usepackage{epsfig}
\usepackage{slashbox}
\usepackage{xcolor}
\usepackage{hyperref}
\usepackage{cite}

\makeatletter \@addtoreset{equation}{section} \makeatother

\def\IZ{\mathbb{Z}}

\def\CA{{\cal A}}
\def\CE{{\cal E}}\def\CF{{\cal F}}
\def\CH{{\cal H}}
\def\CL{{\cal L}}\def\CM{{\cal M}}
\def\CN{{\cal N}}\def\CO{{\cal O}}

\def\a{\alpha}\def\b{\beta}\def\g{\gamma}
\def\d{\delta}\def\e{\epsilon}
\def\th{\theta}
\def\l{\lambda}
\def\m{\mu}\def\n{\nu}
\def\r{\rho}\def\s{\sigma}
\def\t{\tau}
\def\w{\omega}

\def\O{\Omega}

\begin{document}


\begin{titlepage}
\vfill
\begin{flushright}
{\tt\normalsize EFI-15-26}\\
\end{flushright}
\vfill
\begin{center}
{\Large\bf Three-Charge Black Holes and
\vspace*{0.3cm}\\
Quarter BPS States in Little String Theory
}

\vfill
Amit Giveon$^\diamondsuit$,
Jeffrey Harvey$^\clubsuit$,
David Kutasov$^\clubsuit$,
and Sungjay Lee$^\clubsuit$

\vskip 7mm
$^\diamondsuit${\it Racah Institute of Physics, The Hebrew University
\\Jerusalem, 91904, Israel}
\vskip 1mm
$^\clubsuit${\it Enrico Fermi Institute and Department of Physics,
The University of Chicago\\
5620 S. Ellis Av., Chicago, Illinois 60637, USA}

\end{center}
\vfill

\begin{abstract}
\noindent
We show that the system of $k$ NS5-branes wrapping $\mathbb{T}^4\times S^1$ has non-trivial vacuum structure. Different vacua have different spectra of $1/4$ BPS states that carry momentum and winding around the $S^1$. In one vacuum, such states are described by black holes; in another, they can be thought of as perturbative BPS states in Double Scaled Little String Theory. In general, both kinds of states are present. We compute the degeneracy of perturbative BPS states exactly, and show that it differs from that of the corresponding black holes. We comment on the implication of our results to the black hole microstate program, UV/IR mixing in Little String Theory,
string thermodynamics, the string/black hole transition, and other issues.

\end{abstract}

\vfill
\end{titlepage}

\parskip 0.1 cm
\tableofcontents
\newpage
\renewcommand{\thefootnote}{\#\arabic{footnote}}
\setcounter{footnote}{0}

\parskip 0.2 cm


\section{Introduction}

In this paper we will study the system of $k$ NS5-branes
in type II string theory. We will take the fivebranes to wrap $\mathbb{R}^4\times S^1$,
and focus on states that carry momentum $P$ and winding $W$ around the $S^1$.
For general $(P,W)$, the lowest lying states with these quantum numbers preserve
four of the sixteen supercharges preserved by the fivebranes.
Thus, they can be thought of as quarter BPS states in the fivebrane theory.

The states in question have the same quantum numbers as the three-charge
black holes that were studied extensively in the last twenty years in the context
of providing a microscopic interpretation of black hole entropy, starting with the work
of \cite{Strominger:1996sh}; for a recent review see \cite{Mandal:2010cj}.
They also figure prominently in the fuzzball program
that attempts to describe these microstates by horizonless geometries \cite{Mathur:2005zp,Bena:2013dka}.
In these cases, one needs to replace the  $\mathbb{R}^4$ that the
fivebranes are wrapping by a compact manifold, such as $\mathbb{T}^4$,
and we will discuss this case as well.

Our main interest will be in the dependence of the spectrum of
the above states on the positions of the fivebranes.
We will see that it is qualitatively different when the
fivebranes are separated by any finite distance, and when
they are coincident. The two cases are separated by a string-black hole transition.
This may seem surprising, since separating the fivebranes corresponds
in the low energy theory to Higgsing a non-abelian gauge group,
and one would expect that if the W-boson mass scale is low,
the physics of high mass states, such as the ones we will study,
should not be affected. We will discuss why it nevertheless happens,
and comment on some implications.

In the context of black hole physics, the above system is usually discussed in the full,
asymptotically flat space transverse to the fivebranes.
However, one can also study it in the theory obtained by
restricting to the near-horizon geometry of the fivebranes.
This theory is known as Little String Theory (LST).
As we review in section 2, it can be alternatively defined by taking a certain scaling limit
of the full string theory.

In the near-horizon geometry of the coincident fivebranes,
the three-charge black holes are described by certain BPS black hole solutions
in an asymptotically linear dilaton spacetime, which carry the charges $(P,W)$; see e.g.
\cite{Giveon:2005mi,Giveon:2005jv} and references therein.
The entropy of these black holes is given by the familiar result (see section 6)
\begin{align}
  S_\text{BH}=2\pi\sqrt{k PW}\ .\label{ssbbhh}
\end{align}
On the other hand, if one separates the fivebranes in the
transverse $\mathbb{R}^4$, one can study these states
as conventional perturbative string states in a spacetime of the form
$\mathbb{T}^4\times S^1\times \CM_4$, where $\CM_4$ is the background
associated with directions transverse to the fivebranes.
It includes a non-compact direction associated with the radial
direction away from the fivebranes, and some compact directions
associated with the angular part of the geometry.
The precise background depends on the positions of the fivebranes.

If the separations of the fivebranes are sufficiently large,
the string coupling in this background is small everywhere
(unlike the case for coincident fivebranes, where it diverges as one approaches the fivebranes),
and the description of these states as perturbative string states mentioned above is valid.
Thus, one can use standard techniques to count them.

A convenient object for this purpose is the elliptic genus of the
worldsheet CFT corresponding to $\CM_4$. We review the definition of this object and study
its properties in our case in sections 3 and 4. As we discuss there, it can be written as a power
series in a parameter $q$. The coefficient of $q^N$ is the (graded) number of
BPS states with charges $(P,W)$ satisfying $PW=N$. These states are standard perturbative BPS
states  \cite{Dabholkar:1989jt}, for which the right-movers on the worldsheet are
in the ground state while the left-movers are in a general excited state.
Thus, they satisfy
\begin{gather}
 N_R=0\ , ~ N_L=N=PW\ ,
 \nonumber \\
 M= \left|{P\over R}+{WR\over\alpha'}\right|\ , \label{bpsmass}
\end{gather}
where $N_L$ and $N_R$ are the excitation levels for left and right-movers on the worldsheet, $R$ is the radius of the circle the fivebranes wrap, and $M$ is the mass of the BPS state.

We use the elliptic genus to calculate the entropy of perturbative string
states with the same quantum numbers as the black holes mentioned above,
and find the result (for large $PW$)
\begin{align}
  S_\text{string}=2\pi\sqrt{ \Big( 2-\frac1k \Big) PW}\ .
\end{align}
This does not agree with the entropy of black holes with the
same quantum numbers (\ref{ssbbhh}).\footnote{Although the result $S_\text{string}$
is derived when the separations between the fivebranes are
sufficiently large that the string coupling is small everywhere,
we argue in section 5 that it is actually valid whenever the
fivebranes are separated by any finite amount.}
We argue that the system exhibits a phase transition:
when the fivebranes are coincident, quarter BPS states with the
quantum numbers $(P,W)$ correspond to black holes, while
when they are separated they correspond to fundamental strings.
This phenomenon is an example of UV-IR mixing in LST -- turning
on a small IR scale (the masses of W-bosons corresponding to
the separations between the fivebranes) has a large effect on the
spectrum of massive states (the quarter BPS states discussed above).
This UV-IR mixing is possible due to the fact that LST is not a local QFT.

As mentioned above, the black hole point of view requires us to
compactify the worldvolume of the NS5-branes from
$\mathbb{R}^4\times S^1$ to, say $\mathbb{T}^4\times S^1$.
In the compact case, the theory on NS5-branes becomes $(0+1)$
dimensional (i.e. it becomes quantum mechanics). In this case,
the positions of NS5-branes are no longer well defined; instead,
the ground state corresponds to a wavefunction on the classical moduli space.
The transition mentioned above has a slightly different flavor
in this case -- the quantum theory of NS5-branes has non-trivial vacuum structure.
In one vacuum, the fivebranes are coindicent and the entropy of
BPS states is given by $S_\text{BH}$, while in another they are separated
and the entropy is given by $S_\text{string}$.
The UV-IR mixing manifests itself in this theory as the fact
that although the vacuum wavefunction in the string phase has support
in the region where the fivebranes are arbitrarily close to each other,
this phase is nevertheless distinct from the black hole phase,
in which the fivebranes are all coincident. The two phases differ in their UV behavior.

We next turn to the development of the picture presented above
in more detail, starting from the definition and properties of
LST and in particular its holographic description.


\section{Little String Theory and its Scaling Limit}

\subsection{Little String Theory}
The dynamical degrees of freedom localized on NS fivebranes can be decoupled from
bulk degrees of freedom by taking a limit in which the string coupling
$g_s \rightarrow 0$, with the energy scale $E$ held fixed relative to the string
scale, $E \sim m_s$. The resulting theory, known as Little String Theory,
is an interacting six-dimensional theory which does not include gravity,
but otherwise shares many similarities to string theory in asymptotically flat
spacetime, including a Hagedorn density of states\footnote{String theory in
asymptotically flat spacetime exhibits a Hagedorn density of states in an intermediate
energy regime, while at asymptotically high energies the entropy grows faster with the energy.
In LST, the Hagedorn behavior persists up to arbitrarily high energies.}
and T-duality \cite{Berkooz:1997cq,Seiberg:1997zk,Losev:1997hx}.
For reviews see \cite{Aharony:1999ks,Kutasov:2001uf}.

LST has a holographic description in terms of string theory in the
near-horizon geometry of the fivebranes  \cite{Aharony:1998ub}.
For $k$ coincident fivebranes, the near-horizon  metric is given
by \cite{Callan:1991dj,Callan:1991at}
\begin{align}
  ds^2 = dx^\mu dx_\mu + d\phi^2 + 2k d\O_3^2\ ,
\end{align}
where $x^\mu$ ($\mu=0,1,..,5$) parametrize the flat worldvolume
of the fivebranes, $\phi$ is a (function of the) radial coordinate
in the directions transverse to the fivebranes,
and  $d\Omega_3^2$ is the line element on the corresponding angular three-sphere.
The dilaton and $H$-flux in the near-horizon geometry can be written in the form
\begin{align}
  \Phi = -\frac Q2 \phi , \qquad Q = \sqrt{\frac2k}\ ,
\end{align}
and
\begin{align}
  H = 2k \text{Vol}_{S^3}\ ,
\end{align}
where we took $\alpha'=2$.
The four-dimensional space transverse to
the NS5-branes is described by an exactly solvable conformal field
theory \cite{Callan:1991dj,Callan:1991at}, the Callan-Harvey-Strominger (CHS) CFT,
\begin{align}
  \mathbb{R}_\phi \times SU(2)_k \ , \label{cchhss}
\end{align}
that contains a linear dilaton direction with background charge $Q$,
a bosonic $SU(2)$ Wess-Zumino-Witten (WZW) model at level $k-2$
with currents ${\tilde j}^a$ ($a=1,2,3$), and four fermions
$\psi^I = (\psi_a, \psi_\phi)$. The central charge is
\begin{align}
  c = c_\phi + c_\text{SU(2)} +c_\text{f} =
  \big( 1 + 3 Q^2 \big) + \big( 3 - \frac 6k \big) + 2 = 6\ ,
\end{align}
as expected. The total $SU(2)$ currents of the supersymmetric $SU(2)$ WZW model at
level $k$ are given by
\begin{align}
  {\tilde J}^a = {\tilde j}^a - \frac i2 \e^{abc} \psi_b \psi_c \ .
\end{align}
For later convenience we define
\begin{align}
  \psi^\pm = \frac{1}{\sqrt2} \left( \psi_\phi \pm i \psi_3 \right)\ ,
  \qquad
  \tilde \psi^\pm = \frac{1}{\sqrt2} \left( \psi_1 \pm i \psi_2 \right)\ ,
\end{align}
which can be bosonized as
\begin{align}
  \psi^\pm = e^{\pm i H}\ , \qquad
  \tilde \psi^\pm  = e^{\pm i {\tilde H}}\ .
\end{align}

It will be useful in our later analysis to note that one can also describe
the supersymmetric $SU(2)_k$ WZW model as a $\mathbb{Z}_k$ orbifold of the
tensor product of the supersymmetric $U(1)_k$ WZW model and a coset CFT, $SU(2)_k/U(1)$,
\begin{align}
  SU(2)_k \simeq \left( U(1)_k \times \frac{SU(2)_k}{U(1)} \right)/{\mathbb{Z}_k}\ .
  \label{orbifold1}
\end{align}
In this description,  the compact boson $Y$ of the $U(1)_k$ WZW model is
related to the current $\tilde J^3$ via
\begin{align}
  {\tilde J}^3 = {\tilde j}^3 + \psi^+ \psi^- = i \sqrt{\frac k2}  \partial Y\ .
\end{align}
The $SU(2)_k/U(1)$ coset is equivalent to an $N=2$ minimal model whose
central charge is $c=3-{6\over k}$.

The CHS conformal field theory has $\CN=4$ superconformal symmetry
with the superconformal generators
\begin{align}
  G_I = i \left( \psi^I \partial \phi + Q \partial \psi^I \right)
  - Q \eta^a_{IJ} j^a \psi^J + \frac i6 Q \e_{IJKL} \psi^J \psi^K \psi^L \ ,
\end{align}
and the $SU(2)_R$  currents at level one
\begin{align}
  J_\text{R}^a = -\frac i2 {\bar \eta}^a_{IJ} \psi^I \psi^J\ .
\end{align}
Here the 't Hooft tensors $\eta^a_{\m\n}$ and ${\bar \eta}^a_{\m\n}$ are
antisymmetric in $(\m,\n)$ and
construct the Lie algebra of $SU(2)$ from self-dual and anti-self-dual
combinations of $SO(4)$ generators. They are defined explicitly  as
\begin{align}
  \eta^a_{bc} = \e_{abc}\ , & ~~
  \eta^a_{b4} = \delta^a_b\ ,
  \nonumber \\
  {\bar \eta}^a_{bc} = \e_{abc}\ , & ~~
  {\bar \eta}^a_{b4} = -\delta^a_b\ .
\end{align}
In particular, the current $J^3_\text{R}$ is given by
\begin{align}
  2 J^3_\text{R} = \psi^+ \psi^- + \tilde \psi^+ \tilde \psi^-
  = i \partial H + i\partial \tilde H\ .
\end{align}
The normalizable primary vertex operators of the CHS CFT
can be expressed as follows:
\begin{align}
  \CO = e^{-Q(j+1)\phi} e^{i ( \a H + \bar \a \bar H)}
  e^{i (\b \tilde H + \bar \b \bar{\tilde H})}
  \Phi^\text{su}_{\tilde j,\tilde m,\bar{\tilde m}}\ ,
  \label{DSLSTop}
\end{align}
where $\Phi^\text{su}_{\tilde j;\tilde m,\bar{\tilde m}}$ are primary
operators of the bosonic $SU(2)_{k-2}$ current algebra.
The conformal dimensions and R-charges of $\CO$ (\ref{DSLSTop}) are given by
\begin{align}
  h & = - \frac{j(j+1)}{k} + \frac{\tilde j (\tilde j+1)}{k} + \frac{\a^2 + \b^2}{2}\ ,
  \nonumber \\
  r & = \a+\b\ .
\end{align}
In terms of the decomposition $\left( U(1)_k \times \frac{SU(2)_k}{U(1)} \right)/{\mathbb{Z}_k}$,
one can rewrite the contribution of the supersymmetric $SU(2)_k$ WZW model to (\ref{DSLSTop}) as
\begin{align}
  e^{i\b\tilde H} e^{i\bar \b \bar{\tilde H}} \Phi_{\tilde j; \tilde m,\bar{\tilde m}}^\text{su} =
  e^{i\sqrt{\frac{2}{k}}\left[ (\tilde m + \b) Y + (\bar{\tilde m}+\bar \b) \bar Y \right] }
  {\tilde V}^{\text{susy}}_{\tilde j;\tilde m, \bar{\tilde m}}(\b,\bar \b)\ ,
\end{align}
where $\tilde V^{\text{susy}}_{\tilde j;\tilde m, \bar{\tilde m}}(\b,\bar \b)$ denotes
primary operators of the $\CN=2$ minimal model. Their conformal
weights are
\begin{align}
  h = \frac{\tilde j (\tilde j + 1)- (\tilde m+\b)^2}{k} + \frac{\b^2}{2}\ .
\end{align}
As explained in Appendix A, the two parameters $\beta$ and $\bar \b$ can be understood
as spectral flow parameters in the supersymmetric minimal model.

In the CHS background (\ref{cchhss}), the string coupling varies with the distance
from the fivebranes as follows:
\begin{align}
 g_s^2\simeq e^{-Q\phi}\ .\label{strdiv}
\end{align}
Thus $g_s\to 0$ at large distance $(\phi\to\infty)$. This is the boundary of the near-horizon
geometry, analogous to the boundary of AdS for gauge/gravity duality.
At the same time, as one approaches the fivebranes $(\phi\to-\infty)$,
the string coupling diverges. Hence, the exact background (\ref{cchhss})
is not useful for worldsheet calculations, which rely on weak coupling.
To make it useful, we need to do something about the strong coupling region.
One way to deal with it is described in the next subsection.

\subsection{Double Scaling Limit}

From the discussion in the previous subsection, it is clear
that the CHS geometry is applicable only for $k\ge2$ (coincident) fivebranes,
since otherwise the bosonic $SU(2)$ WZW model at level $k-2$ does not make sense.
Thus, fundamental strings propagating in the vicinity of a single NS5-brane
do not see a CHS throat geometry (\ref{cchhss}). As a consequence, in the near-horizon
geometry of $k$ separated fivebranes, the string coupling is bounded from
above \cite{Giveon:1999px,Giveon:1999tq}.
The maximal value of $g_s$ depends on the separations of the fivebranes.

One can arrange the separations such that the coupling is small everywhere.
This amounts to demanding that the masses of
D-strings stretched between different NS5-branes in type IIB string
theory\footnote{The IIA case is very similar.}, denoted collectively by $M_W$,
are much larger than the string scale,
\begin{align}
  M_W \gg m_s \ .
\end{align}
As we review below, the resulting theory can be studied
using perturbative string techniques where $m_s/M_W$ plays the role of the string coupling.
This theory is known as Double Scaled Little String Theory
(DSLST) \cite{Giveon:1999px,Giveon:1999tq}.

\begin{figure}
\centering
\includegraphics[width=0.3\textwidth]{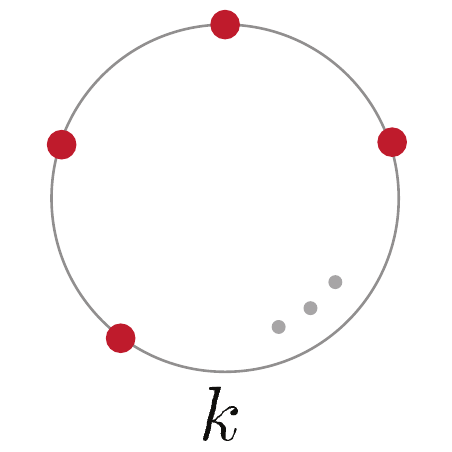}
\caption{NS5-branes on a circle.}\label{figure1}
\end{figure}
For example, consider a configuration of fivebranes arranged
equidistantely on a circle of radius $R_0$ in $\mathbb{R}^4$, depicted in figure \ref{figure1}.
In this configuration, the $SU(k)$ gauge symmetry on the fivebranes is broken to
$U(1)^{k-1}$ at the scale $M_W\sim R_0/g_s l_s^2$. To study the dynamics of the
fivebranes in this case, we can take the double scaling limit
\begin{align}
  g_s \to 0 ~,~  \frac{R_0}{l_s} \to 0
  ~\text{ with }~ \frac{R_0}{g_s l_s} ~\text{fixed.}\nonumber\
\end{align}
Keeping the dimensionless constant $\frac{R_0}{g_s l_s}$ fixed means
keeping the masses of W-bosons fixed in string units.
If these masses are large relative to $m_s$, the theory is weakly coupled
and can be studied using worldsheet techniques; it is a special case of the
DSLST construction mentioned above.

While the DSLST construction is general, the configuration of figure \ref{figure1}
is special in that the corresponding worldsheet CFT is solvable,
which is not the case for generic points in the moduli space of DSLST.
The CHS background (\ref{cchhss}) is replaced in this case by
\begin{align}
  \mathbb{R}_\phi \times
  \left( U(1)_k \times \frac{SU(2)_k}{U(1)} \right)/{\mathbb{Z}_k}
  \ \to \
  \left( \frac{SL(2,\mathbb{R})_k}{U(1)} \times \frac{SU(2)_k}{U(1)} \right)/{\mathbb{Z}_k}\ .
  \label{temp1}
\end{align}
The coset $\frac{SL(2,\mathbb{R})_k}{U(1)}$ describes a $\sigma$-model
on a cigar with asymptotically linear dilaton.
The string coupling grows towards the tip of the cigar,
and attains its largest value at the tip,
\begin{align}
  g_s(\text{tip}) \simeq \frac{m_s}{M_W} \simeq \frac{g_s l_s}{R_0}\label{pwpwpw}
\end{align}
One way to understand this relation is to note that D-strings
stretched between the fivebranes correspond in the deformed theory (\ref{temp1})
to D0-branes located at the tip of the cigar (and particular boundary states
in the $\CN=2$ minimal model). Note that (\ref{pwpwpw})
has the property that as $R_0$ increases, the maximal value of the string
coupling decreases. On the other hand, as $R_0\to 0$ the string coupling at
the tip of the cigar grows without bound. These properties are in agreement
with the expectations mentioned above -- the length of the fivebrane throat
increases with decreasing $R_0$ and vice versa.

The relation between the CHS background (\ref{cchhss}) corresponding to coincident fivebranes
and the background corresponding to fivebranes on a circle (the r.h.s. of (\ref{temp1}))
can be understood by looking at the region far from the tip of the cigar.
We will refer to it below as the CHS region. In that region,
the cigar geometry reduces to a cylinder $ \mathbb{R}_\phi \times S^1$,
and the CFT on the r.h.s. of (\ref{temp1}) reduces to that on the l.h.s. which,
as explained above, is equivalent to (\ref{cchhss}).
This agrees with the intuition that far from the fivebranes one does not notice
that they are separated, and the background of separated fivebranes
should reduce to that of coincident ones.

Both the $SL(2,\mathbb{R})/U(1)$  and the $SU(2)/U(1)$  CFT's have $\CN=2$ superconformal
symmetry. We will denote their $U(1)_R$ currents by $J_\text{R}^\text{sl}$ and
$J_\text{R}^\text{su}$, respectively. Some  basic properties of these two coset
models are summarized in Appendix A. While the tensor product of cigar and
minimal models preserves $\CN=2$ superconformal symmetry, one can show that
the $\mathbb{Z}_k$ orbifold in (\ref{temp1}) enhances the superconformal symmetry
to $\CN=4$, in agreement with the fact that the background (\ref{temp1}) can be thought of
as describing a near-singular non-compact $K3$ surface.

The CSA of the $SU(2)_R$ current that belongs to the $\CN=4$ algebra can be taken to be
\begin{align}
  2J^3_\text{R} =  J_\text{R}^\text{sl} + J_\text{R}^\text{su}\ .
\end{align}
The factor of two on the l.h.s. is due to the fact that $\CN=4$ supercharges transform
as a doublet under $SU(2)_R$, and hence naturally have  $U(1)_R$ charge $\pm1/2$,
while the $U(1)_R$ current in the $\CN=2$ algebra is usually normalized so that the
supercharges carry charge $\pm 1$.

Finally we discuss the normalizable vertex operators in the CFT (\ref{temp1}).
The primary operators of the cigar CFT, $V_{j;m,\bar m}^\text{susy}(\a,\bar \a)$,
are defined and discussed in Appendix A. They have conformal weight and $U(1)_R$ charge
\begin{align}
  h & = \frac{(m+\a)^2-j(j+1)}{k} + \frac{\a^2}{2}\ ,
  \nonumber \\
  r & = \frac{2(m+\a)}{k} + \a \ .
\end{align}
In the CHS region, the vertex operator $V^\text{susy}_{j;m,\bar m}(\a,\bar \a)$ behaves like
\begin{align}
  V^\text{susy}_{j;m,\bar m}(\a,\bar \a) \ \simeq \
  e^{-Q (j+1) \phi} e^{i\a H} e^{i \bar \a \bar H}
  e^{i \sqrt{\frac 2k}\left[ (m+\a) Y + (\bar m+\bar \a) \bar Y \right]}\ .
\end{align}
Note that the parameters $\a$ and $\bar \a$ can be identified as
spectral flow parameters.

The primary operators of the minimal model
$\tilde V^\text{susy}_{\tilde j;\tilde m, \bar{\tilde m}}(\b,\bar \b)$,
are similarly defined and discussed in Appendix A.
Their conformal weight and $U(1)_R$ charge are
\begin{align}
  h & = \frac{\tilde j ( \tilde j + 1) - (\tilde m+\b)^2}{k} + \frac{\b^2}{2}\ ,
  \nonumber \\
  r & = -\frac{2(\tilde m+\b)}{k} + \b
\end{align}
In the CHS region one has
\begin{align}
  e^{i\sqrt{\frac{2}{k}}\left[ (\tilde m + \b) Y + (\bar{\tilde m}+\bar \b) \bar Y \right] }
  {\tilde V}^{\text{susy}}_{\tilde j;\tilde m, \bar{\tilde m}}(\b,\bar \b)
  \simeq
  e^{i\b\tilde H} e^{i\bar \b \bar{\tilde H}} \Phi_{\tilde j; \tilde m,\bar{\tilde m}}^\text{su}\ .
\end{align}
This can be used to construct a normalizable vertex operator in the full
DSLST background (\ref{temp1}), that is asymptotic to the vertex operator (\ref{DSLSTop})
in the CHS region,
\begin{align}
  V_{j;m,\bar m}^\text{susy}(\a,\bar \a) \cdot
  {\tilde V}^\text{susy}_{\tilde j;\tilde m, \bar{\tilde m}}(\b,\bar \b)
  \ \to \
  e^{-Q(j+1)\phi} e^{i ( \a H + \bar \a \bar H)} e^{i (\b \tilde H + \bar \b \bar{\tilde H})}
  \Phi^\text{su}_{\tilde j,\tilde m,\bar{\tilde m}}
\end{align}
with
\begin{align}
  m + \a & = \tilde m + \b \ ,
  \nonumber \\
  \bar m + \bar \a & = \bar{\tilde m} + \bar \b\ .
\end{align}
%


\section{Elliptic Genus of DSLST}

The holographic worldsheet description of DSLST at a generic point
in its moduli space involves the background $\CM^{5,1}\times \CM_4$,
where $\CM_4$ $(\CM^{5,1})$ is a CFT associated with the directions
transverse to (along) the fivebranes. A natural quantity to consider is
the elliptic genus of the CFT on $\CM_4$, defined as
\begin{align}
  \CE_\text{DSLST} =\text{Tr}_{\CH_\text{RR}}\Big[ (-1)^F q^{L_0-\frac{c}{24}}
  {\bar q}^{\bar L_0 - \frac{\bar c}{24}} e^{2\pi i z (2J_\text{R}^3)_0}\Big]\ ,
  \label{ellgen}
\end{align}
where $q=e^{2\pi i \t}$, $c=\bar c = 6$ are the left and
right-moving central charges, and $J_\text{R}^3$
is a Cartan generator of the left-moving $SU(2)_R$ symmetry.
The trace is taken over states in the Ramond-Ramond sector of the SCFT.
We discussed the spacetime significance of this quantity in the introduction.
In this section, we will calculate it using worldsheet techniques.
In particular the calculation will be performed at the point in moduli space corresponding
to figure \ref{figure1}, where
\begin{align}
\CM_4=\left( \frac{SL(2,\mathbb{R})_k}{U(1)} \times \frac{SU(2)_k}{U(1)} \right)/{\mathbb{Z}_k}\  \label{dslstt}
\end{align}
but, as we will discuss later, the result is independent of the moduli.

We will see that the elliptic genus is not holomorphic in $q$ unlike the situation
for compact SCFT's. This is because both discrete and continuum
states can contribute to the elliptic genus
\cite{Troost:2010ud, Eguchi:2010cb, Ashok:2011cy, Harvey:2014nha}.
We will discuss a physical way to separate the elliptic genus into two contributions, corresponding to the
discrete and continuum states, respectively.

Some properties of the elliptic genus of DSLST will be  discussed in the next section.

\subsection{Cigar CFT}

We can describe the cigar CFT as a two dimensional $\CN=(2,2)$ non-linear $\sigma$-model
whose Lagrangian takes the form
\begin{align}
  \CL = - g_{i\bar j} \partial^\mu \phi^i \partial_\mu {\bar \phi}^{\bar j} + i
  g_{i\bar j} {\bar \psi}^{\bar j}_- D_+ \psi_-^i +
  i g_{i\bar j} {\bar \psi}^{\bar j}_+ D_- \psi_+^i +
  R_{i\bar j k \bar l} \psi^i_+ \psi_-^k {\bar \psi}^{\bar j}_- {\bar \psi}^{\bar l}_+\ ,
\end{align}
where the target space metric is
\begin{align}
  ds^2 = k \left( d r^2 + \tanh^2 r d\theta^2 \right) \ .
\end{align}
The non-linear $\sigma$ model also includes a non-trivial dilaton profile, which will not play a role in what follows. The four global supercharges are given by
\begin{align}
  Q_\pm = \int d\s \ 2 g_{i\bar j} \partial_\pm \bar{\phi}^{\bar j} \psi_\pm^i\ ,
  \qquad
  \bar Q_\pm = \int d\s \ 2 g_{i\bar j}  {\bar \psi}_\pm^{\bar j} \partial_\pm {\phi}^{i}\ .
\end{align}
Using supersymmetric localization, one can reduce the path-integral of the cigar CFT to a finite dimensional integral over the holonomy-torus \cite{Murthy:2013mya, Ashok:2013pya},
\begin{align}
  \CE_\text{cig}(\t,z) & = k \int_0^1 ds_1 \int_0^1 ds_2
  \sum_{(n,w)\in \mathbb{Z}^2}
  \frac{\vartheta_{1}\big(\t, s_1\t+s_2+z(1+1/k)\big)}
  {\vartheta_{1}\big(\t, s_1\t+s_2+z\big)}
  \nonumber \\ & \times
  e^{-2\pi i z w}
  e^{-\frac{\pi k}{\t_2}\big| s_1\t+s_2 + n + \t\w\big|^2}\ ,
  \label{result1againrep}
\end{align}
where $\vartheta_1(\t,z)$ denotes the odd Jacobi theta function given by
\begin{align}
  \vartheta_1(\tau,z) = -iq^{1/8} e^{\pi i z} \prod_{n=1}^\infty
  (1-q^n) (1- e^{2\pi i z} q^n) (1- e^{-2\pi i z} q^{n-1})\ .
\end{align}
Using the Poisson resummation formula, one can rewrite the elliptic genus as
\begin{align}
  \CE_\text{cig}(\t,z) & = \sqrt{k\t_2} \int_0^1 ds_1 \int_0^1 ds_2
  \sum_{(p,w)\in \mathbb{Z}^2}
  \frac{\vartheta_{1}\big(\t, s_1\t+s_2+z(1+1/k)\big)}
  {\vartheta_{1}\big(\t, s_1\t+s_2+z\big)}
  \nonumber \\ & \times
  e^{-2\pi i z w} e^{-2\pi i s_2 p} q^{l_0} {\bar q}^{\bar l_0}\ ,
  \label{6.1}
\end{align}
where
\begin{align}
  l_0 = \frac{1}{4k} \big( p- k(w+u_2) \big)^2 \ ,
  \qquad
  \bar l_0 = \frac{1}{4k} \big( p + k(w+u_2) \big)^2\ .
\end{align}
The integers $(p,w)$ are the momentum and winding around the cigar.

Following the treatment in
\cite{Troost:2010ud, Eguchi:2010cb, Murthy:2013mya, Ashok:2013pya},
one can show that
\begin{align}
  \CE_\text{cig} = \CE_\text{d} + \CE_\text{c}\ ,
\end{align}
where the contribution from the discrete spectrum is
\begin{align}
  \CE_d & = +\frac{i\vartheta_{1}(\t,z) }{\eta(q)^3}
  \sum_{\a=0}^{k-1} \sum_{w\in \mathbb{Z}} \frac{q^{(-\a+kw) w} e^{2\pi i z(-\a/k +2w)}}{1- e^{2\pi i z} q^{-\a+kw}}
  \nonumber \\ & =
  + \frac{i\vartheta_{1}(\t,z) }{k \eta(q)^3} \sum_{\b,\g=1}^{k} e^{2\pi i \frac{\b\g}{k}} \sum_{w\in \mathbb{Z}}
  \frac{q^{\frac{(k w + \b)^2}{k}} \left( e^{2 \pi i z} \right)^{2\frac{k w + \b}{k}}}
  {1- \left( e^{2\pi i z} \right)^{\frac1k} q^{\frac{kw +\b}{k}} e^{2\pi i \frac{\g}{k}}}
  \label{resultaaa}
\end{align}
and that of the scattering states is
\begin{align}
  \CE_\text{c} & = -
  \frac{\vartheta_{1}(\t,z)}{2\eta(q)^3} \sum_{p,w\in \mathbb{Z}}
  \int_{-\infty}^{\infty} dK  \ \frac{1}{\pi}\left[ \frac{1}{K +i (p+kw)} \right]
  \left( q\bar q \right)^{\frac{K^2}{4k}}
  q^{\frac{1}{4k}\left(p - k w \right)^2}
  {\bar q}^{\frac{1}{4k}\left(p + k w \right)^2} e^{2\pi i z (\frac{p}{k}-w)}
  \nonumber \\ & =
  - \frac{\vartheta_{1}(\t,z)}{2\eta(q)^3} \sum_{p,w\in \mathbb{Z}}
  \int_{0}^{\infty} dK  \ \frac{1}{\pi}\left[ \frac{1}{K +i (p+kw)} +\frac{1}{-K + i(p+kw)} \right]
  \left( q\bar q \right)^{\frac{K^2}{4k}}
  \nonumber \\ & \times
  q^{\frac{1}{4k}\left(p - k w \right)^2}
  {\bar q}^{\frac{1}{4k}\left(p + k w \right)^2} e^{2\pi i z (\frac{p}{k}-w)}
  \label{result4}
\end{align}
When $p+kw=0$, we choose the integration contour in (\ref{result4}) slightly above
the real axis.
The discrete contribution to the elliptic genus is holomorphic in $q$ but not modular, while the
contribution from the scattering states restores modularity at the cost of a loss of holomorphy.
Note that the scattering state contribution to the elliptic genus at $z=0$, i.e. to the Witten index,
vanishes.

Below we discuss a more physical way to compute the contribution of the scattering states
to the elliptic genus.  This method should be applicable to any non-compact CFT,
and in particular to DSLST at a generic point in its moduli
space\footnote{The universality of this contribution played an important role in \cite{Giveon:2014hfa}.}.
The reader not interested in the details can skip section 3.1.1 and proceed directly to section 3.1.2.

\subsubsection{Spectral asymmetry and non-holomorphic contributions}

The contribution of the continuum of $\delta$-function normalizable states
to the elliptic genus is related to the difference between the densities
of bosonic and fermionic states.
These densities can be computed  from the corresponding scattering phase shifts.
The individual phase shifts for bosons and fermions are non-trivial,
but the difference between them can be computed exactly using only asymptotic data.
The general idea goes back to calculations of the Witten index in
supersymmetric quantum mechanics with non-compact target space
\cite{Akhoury:1984pt,Ashok:2014nua,Pioline:2015wza}.

In order to perform this computation, we first consider the Scherk-Schwarz
reduction of the cigar $\sigma$-model in a sector with winding number $w$ to
quantum mechanics, i.e. we take
\begin{align}
  \partial_1 r = 0 \ , \qquad \partial_1 \theta = w\ .
\end{align}
The bosonic part of the resulting Lagrangian takes the form
\begin{align}
  \CL_\text{QM}^B = \frac{k}{2} \left( \frac{d  r}{dt}\right)^2
  + \frac k2 \tanh^2{r} \left( \frac{d \theta}{dt} \right)^2
  - \frac{k}{2} w^2 \tanh^2{r} \ .
  \label{CigarQM1a}
\end{align}
This Lagrangian describes the center of mass motion of a string
winding the cigar $w$ times.  The attractive potential, the last term in (\ref{CigarQM1a}),
is due to the fact that the string can decrease its energy by moving towards
the tip of the cigar.

Quantizing the system, the fermion operators have to satisfy the
canonical anticommutation relations
\begin{align}
  \big\{ \bar \psi_+, \psi_+ \big\} =\big\{  \bar \psi_-, \psi_- \big\} = \frac{1}{k}\ ,
\end{align}
where $\psi_\pm = e_m \psi_\pm^m$ ($m=1,2$) with an orthonormal frame $e_m$. They can be
represented by four-dimensional Dirac gamma matrices
\begin{gather}
  \sqrt k \bar \psi_+  =  \frac{\s^1 + i \s^2}{2} \otimes {\bf 1}_2 \ , \qquad
  \sqrt k \psi_+ = \frac{\s^1 - i \s^2}{2} \otimes {\bf 1}_2 \ ,
  \nonumber \\
  \sqrt k \bar \psi_-  = \s^3 \otimes \frac{\s^1 + i \s^2}{2}\ , \qquad
  \sqrt k \psi_-  = \s^3 \otimes \frac{\s^1 - i \s^2}{2}\ ,
\end{gather}
where $\vec \s$ represent the Pauli matrices. The fermion number operator then becomes
\begin{align}
  F = \frac{1 - \s^3}{2} \otimes {\bf 1}_2 + {\bf 1}_2 \otimes \frac{1 - \s^3}{2}\ .
\end{align}
One can regard the wavefunctions as 4-component spinors, two of which are bosonic and the others
are fermionic, i.e.,
\begin{align}
  \langle  x | B_1 \rangle & = f_1(r,\th) |++\rangle \ ,
  \nonumber \\
  \langle  x | F_1 \rangle & = g_1(r,\th) |-+\rangle \ ,
  \nonumber \\
  \langle  x | B_2 \rangle & = f_2(r,\th) |--\rangle \ ,
  \nonumber \\
  \langle  x | F_2 \rangle & = g_2(r,\th) |+-\rangle \ .
  \label{wave1}
\end{align}
Their left and right-moving $U(1)_R$ charges denoted by
$U(1)_l$ and $U(1)_r$ respectively are summarized, up to overall constants,
in the table below
\begin{center}
\setlength{\tabcolsep}{10pt}{
\begin{tabular}{c|cccc}
  \hline
  & $|B_1\rangle$ & $|F_1\rangle$ & $|B_2\rangle$ & $|F_2\rangle$
  \\ \hline
  $U(1)_l$ & $- \frac12 $ & $-\frac12$ & $+\frac12$ & $+\frac12$
  \\
  $U(1)_r$ & $- \frac12 $ & $+\frac12$ & $+\frac12$ & $-\frac12$
\end{tabular}\ .
}
\end{center}

For the scattering states,
the asymptotic behavior of $f_{1,2}(r,\th)$ and $g_{1,2}(r,\th)$ in (\ref{wave1}) can be
described as follows
\begin{align}
  f_1(r,\th) \ & \to \ \left( e^{-iKr} + e^{2i\d_B^1(K)} e^{+iKr} \right)\cdot e^{ip\th}\ ,
  \nonumber \\
  g_1(r,\th) \ & \to \ \left( e^{-iKr} + e^{2i\d_F^1(K)} e^{+iKr} \right)\cdot e^{ip\th}\ ,
  \nonumber \\
  f_2(r,\th) \ & \to \ \left( e^{-iKr} + e^{2i\d_B^2(K)} e^{+iKr} \right)\cdot e^{ip\th}\ ,
  \nonumber \\
  g_2(r,\th) \ & \to \ \left( e^{-iKr} + e^{2i\d_F^2(K)} e^{+iKr} \right)\cdot e^{ip\th}\ .
\end{align}
The boson and fermion scattering states are paired by the supercharges $Q_+$ and $\bar Q_+$.
More precisely,
\begin{align}
  Q_+ |B_1 \rangle \sim |F_1 \rangle\ , \qquad
  Q_+ |F_2 \rangle \sim |B_2 \rangle\ ,
\end{align}
which implies that, in the limit $r\to \infty$,
\begin{align}
  \left[ - \frac{i}{\sqrt k} \frac{\partial}{\partial r} - \frac{1}{\sqrt k}
  \frac{\partial}{\partial \th} -i \sqrt k w \right] f_1(r,\th) & \propto g_1(r,\th)\ ,
  \nonumber \\
  \left[ - \frac{i}{\sqrt k} \frac{\partial}{\partial r} - \frac{1}{\sqrt k}
  \frac{\partial}{\partial \th} -i \sqrt k w \right] g_2(r,\th) & \propto f_2(r,\th)\ .
\end{align}
These relations provide very strong constraints between phase-shift factors,
\begin{align}
  e^{2i \big(\d_B^1(K) - \d_F^1(K)\big)} & = - \frac{K+i(p+kw)}{K-i(p+kw)}\ ,
  \nonumber \\
  e^{2i \big( \d_B^2(K) - \d_F^2(K)\big) } & = - \frac{K-i(p+kw)}{K+i(p+kw)}\ .
\end{align}
This result can be verified directly by using the exact results
for the bosonic and fermionic phase shifts \cite{Giveon:2003wn,Ashok:2011cy}.
Note that the individual phase shifts receive non-trivial
stringy corrections that play an important role in the discussion of
\cite{Giveon:2015cma,Ben-Israel:2015mda}.
However, these stringy corrections cancel in
the difference of phase shifts, which is given
exactly by the quantum mechanical result.

Using the standard relation between the spectral density
and the phase shift in quantum mechanics,
\begin{align}
   \rho_B(E) - \rho_F(E) = \frac{1}{\pi}
   \frac{\partial}{\partial E} \left( \d_B(E) - \d_F(E) \right) \ ,
\end{align}
one obtains the difference in the density of states
\begin{align}
  \rho_B^1(K) - \rho_F^1(K) = - \rho_B^2(K) + \rho_F^2(K)  =
  \frac{1}{2\pi i} \left[ \frac{1}{K +i (p+kw)} +\frac{1}{-K + i(p+kw)}\right]\ .
  \label{result1a}
\end{align}
This result implies that, unless one turns on
a chemical potential $z$ for the R-charge,
there is no spectral asymmetry. It explains
why the scattering state contribution to
the Witten index vanish, $\CE_c(\t,z=0)=0$.

In a sector with winding $w$ around the cigar,
the contribution of scattering states to the elliptic genus with
chemical potential $z$ can be expressed as
\begin{align}
  \CE_\text{c}^{QM}(w) & = \sum_{p\in \mathbb{Z}} \int_0^\infty dK \
  \Big[\left( \r_B^1(K) - \r_F^1(K) \right)  e^{-\pi i z}
  + \left( \r_B^2(K) - \r_F^2(K) \right)  e^{+\pi i z} \Big]
  \nonumber\\ & \times
  \left( e^{-2\pi \t_2} \right)^{E(K,p,w)}
  \left( e^{2\pi i \t_1} \right)^{P(p,w)} e^{2\pi i z J(p,w)}\ ,
\end{align}
where
\begin{align}
  E(K,p,w) & = \frac{K^2}{2k} + \frac{p^2}{2k} + \frac{k w^2}{2}\ ,
  \nonumber \\
  P(p,w) & = - pw\ ,
  \nonumber \\
  J(p,w)  & = \frac pk - w\ .
\end{align}
Using the result (\ref{result1a}), one can rewrite the index $\CE_c^{QM}$ in the
following form
\begin{align}
  \CE_c^{QM}(w) & =  - \sin{\pi z} \sum_{p}
  \int_{0}^{\infty} dK  \ \frac{1}{\pi}\left[ \frac{1}{K +i (p+kw)} +\frac{1}{-K + i(p+kw)} \right]
  \left( q\bar q \right)^{\frac{K^2}{4k}}
  \nonumber \\ & \times q^{\frac{1}{4k}\left(p - k w \right)^2}
  {\bar q}^{\frac{1}{4k}\left(p + k w \right)^2} e^{2\pi i z (\frac pk-w)}\ .
\end{align}
Since the difference in the spectral densities
is not affected by the oscillator modes of the string,
the continuum part of the elliptic genus can be written as
\begin{align}
  \CE_c = \CE_c^{osc} \times \CE_c^{com} \ , \qquad \CE_c^{com} = \sum_{w} \CE_c^{QM}(w) \ ,
\end{align}
where the contribution from the oscillator modes is
\begin{align}
  \CE_c^{osc} = \prod_{n=1}^\infty \left[ \frac{\left(1- q^n e^{2\pi i z}\right)
  \left(1- q^n e^{-2\pi i z}\right)}{\left( 1-q^n \right)^2}\right]\ .
\end{align}
Using the definition of the Jacobi theta function $\vartheta_{1}(\t,z)$
and the Dedekind eta function $\eta(q)$,
one can rewrite $\CE_c$ in the  form
\begin{align}
  \CE_\text{c} &= -\frac{\vartheta_{1}(\t,z)}{2\eta(q)^3} \sum_{p,w\in\mathbb{Z}}
  \int_{0}^{\infty} dK  \ \frac{1}{\pi}\left[ \frac{1}{K +i (p+kw)} +\frac{1}{-K + i(p+kw)} \right]
  \left( q\bar q \right)^{\frac{K^2}{4k}}
  \nonumber \\ & \times
   q^{\frac{1}{4k}\left(p - k w \right)^2}
  {\bar q}^{\frac{1}{4k}\left(p + k w \right)^2} e^{2\pi i z (\frac pk-w)}\ ,
  \label{result9}
\end{align}
which agrees with (\ref{result4}).

\subsubsection{Mock modularity}

There is a close relationship between mock modular forms
and the elliptic genera of non-compact CFT's \cite{Troost:2010ud, Eguchi:2010cb}. In fact,
one can show that
the discrete part (\ref{resultaaa}) can be written as follows
\begin{align}
  \CE_d & = + \frac{i\vartheta_{1}(\t,z) }{k\eta(q)^3} \sum_{\b,\g=1}^{k}
  e^{2\pi i \frac{\b\g}{k}} q^{\frac{\b^2}{k}} \left( e^{2\pi i z}\right)^{\frac{2\b}{k}}
  \CA_{1,k}\big(\t, \frac{z+ \b\t +\g}{k}\big)
  \label{resultbbb}\ ,
\end{align}
where the Appell-Lerch sum is a well-known mock modular form defined by \cite{Eguchi:2010cb}
\begin{align}
  \CA_{1,k}(\t,z)= \sum_{t\in\mathbb{Z}}
  \frac{q^{k t^2} \left( e^{2\pi i z} \right)^{2kt}}{1- \left( e^{2\pi i z}\right)q^t}\ .
  \label{ES}
\end{align}
On the other hand, from an integration formula,
\begin{align}
  \int_{\mathbb{R}\mp i\e} dp\  \frac{1}{p -i\l} e^{-\a p^2} = i\pi \text{sgn}(\l\pm\e)
  \text{Erfc}\big(\sqrt\a |\l| \big) e^{\a \l^2} \ \
  \left( \a,\e>0 \text{ and } \l \in \mathbb{R} \right)\ ,
  \label{identity01}
\end{align}
one can express the continuum part (\ref{result4}) as
\begin{align}
  \CE_\text{c} =
  \frac{i\vartheta_{1}(\t,z)}{2 \eta(q)^3} \sum_{l=0}^{2k-1} R^-_{k,l}(\t)
  \vartheta_{k,l}\left(\t,\frac zk\right)\ ,
  \label{continuum1}
\end{align}
where the non-holomorphic Eichler integrals $R^\pm_{k,l}(\t)$ are defined as
\begin{align}
  R^\pm_{k,l}(\t) = \sum_{\l=l+2k\mathbb{Z}} \text{sgn}(\l\pm\e)
  \text{Erfc}\left(\sqrt{\frac{\pi\t_2}{k}} |\l| \right) q^{-\frac{\l^2}{4k}}\ ,
  \label{identity02}
\end{align}
and $\vartheta_{k,l}(\t,z)$ denote Jacobi theta functions at level
$k$
\begin{align}
  \vartheta_{k,l}(\t,z)
  = \sum_{\l=l +2k \mathbb{Z} } q^{\l^2/4k} e^{2\pi i z \l}\ .
\end{align}
Since the Eichler-Zagier involution maps a Jacobi theta function at
level $k$ to a different Jacobi theta function
\begin{align}
  \frac 1k \sum_{\b,\g=1}^{k} e^{2\pi i \frac{\b\g}{k}} q^{\frac{\b^2}{k}}
  \left( e^{2\pi i z} \right)^{2\b} \vartheta_{k,l}\big(\t,z+\frac{\b\t +\g}{k}\big)
  = \vartheta_{k,2k-l}\big(\t,z\big) \ \text{ for } \ l\in \mathbb{Z}_{2k}\ ,
  \label{identity05}
\end{align}
(\ref{continuum1}) can be written in the form
\begin{align}
  \CE_\text{c} =
  \frac{i\vartheta_{1}(\t,z)}{\eta(q)^3}
  \cdot
  \frac1k \sum_{\b,\g=1}^{k} e^{2\pi i \frac{\b\g}{k}} q^{\frac{\b^2}{k}}
  \left( e^{2\pi i \frac zk} \right)^{2\b}
  \cdot
  \left(-\frac12 \sum_{l=1}^{2k} R^+_{k,l}(\t)
  \vartheta_{k,l}\left(\t,\frac{z+\b\t+\g}{k}\right) \right)\ .
\end{align}
Collecting all the results, one can finally observe that
the full elliptic genus can be expressed in terms of the
non-holomorphic modular completion of the Appell-Lerch
sum $\CA_{1,k}(\t,z)$,
\begin{align}
  \CE_\text{cig} =\CE_d+\CE_c  & = + \frac{i\vartheta_{1}(\t,z) }{k\eta(q)^3} \sum_{\b,\g=1}^{k}
  e^{2\pi i \frac{\b\g}{k}} q^{\frac{\b^2}{k}} \left( e^{2\pi i z}\right)^{\frac{2\b}{k}}
  {\hat \CA_{1,k}}\big(\t, \frac{z+ \b\t +\g}{k}\big)\ ,
  \label{resultccc}
\end{align}
where
\begin{align}
  \hat \CA_{1,k}(\t,z) = \CA_{1,k}(\t,z) - \frac12 \sum_{l=1}^{2k} R^+_{k,l}(\t)
  \vartheta_{k,l}(\t,z) \ .
  \label{defa}
\end{align}
Thus, the elliptic genus of the cigar CFT is expressed as the Eichler-Zagier
involution  \cite{EZ} of the modular completion of Appell-Lerch sum, $\hat \CA_{1,k}(\t,z)$.

\subsubsection{Character decomposition}

We now discuss the expansion of the elliptic genus $\CE_\text{cig}$
(\ref{resultaaa}) in terms of $\CN=2$ superconformal characters
in the Ramond sector with an insertion of $(-1)^F$.

Let us first introduce the $\CN=2$ character formula
with $c_\text{cig} =3\left(1+\frac 2k \right)>3$ of
discrete representations \cite{Eguchi:2004yi,Israel:2004jt}
\begin{align}
  \text{Ch}^\text{cig}_{l,n}(\t,z) = q^{\frac{n^2 - (l-1)^2}{4k}}
  \left( e^{2\pi i z} \right)^{\frac{n}{k}}
  \frac{1}{1 - e^{2\pi i z} q^{\frac{n-l+1}{2}}} \cdot
  \frac{i\vartheta_{1}(\t,z)}{\eta(q)^3}\ ,
\end{align}
where $1 \leq l \leq k+1$. The conformal weight $h$ and
the $U(1)_R$ charge $r$ of an $\CN=2$ primary
corresponding to each character are
\begin{enumerate}

\item[$\bullet$] $\boldsymbol{n-l+1\geq 2}$:
  \begin{align}
    h_{l,n} - \frac{c_\text{cig}}{24} & = \frac{n^2 - (l-1)^2}{4k}\ ,
    \nonumber \\
    r_{l,n} & = \frac{n}{k} - \frac12\ .
  \end{align}

\item[$\bullet$] $\boldsymbol{n-l+1\leq -2}$:
  \begin{align}
    h_{l,n} - \frac{c_\text{cig}}{24} & = \frac{(k-n)^2 - (k-l+1)^2}{4k}\ ,
    \nonumber \\
    r_{l,n} & = -\frac{k-n}{k} + \frac12\ .
  \end{align}

\item[$\bullet$] $\boldsymbol{n-l+1 = 0}$:
  \begin{align}
    h_{l,n} - \frac{c_\text{cig}}{24} & = 0\ ,
    \nonumber \\
    r_{l,n} & = \frac{n}{k} - \frac12\ .
  \end{align}

\end{enumerate}
The $\CN=2$ characters enjoy a $\mathbb{Z}_2$ reflection symmetry
\begin{align}
  \text{Ch}^\text{cig}_{l,n}(\t,z) =
  \text{Ch}^\text{cig}_{(k+2)-l,k-n}(\t,-z)\ ,
\end{align}
and transform under the spectral flow by $\a$ units as
\begin{align}
  q^{\frac{c_\text{cig}}{6} \a^2 } \left(e^{2\pi i z}\right)^{\frac{c_\text{cig}}{3} \a}
  \text{Ch}^\text{cig}_{l,n}(\t,z+\t \a)
  = (-1)^\a \text{Ch}^\text{cig}_{l,n+2\a}(\t,z)\ .
\end{align}
It is straightforward to show that the discrete part of the elliptic genus of
the $\frac{SL(2)_k}{U(1)}$ CFT can be expanded as
\begin{align}
  \CE_d = \sum_{\tilde l=0}^{k-1}
  \sum_{w\in \mathbb{Z}}
  \text{Ch}^\text{cig}_{\tilde l+ 1 ,-\tilde l + 2kw}(\t,z )\ .
  \label{cigellch}
\end{align}

\subsection{Minimal model}

The Landau-Ginzburg theory with a superpotential
\begin{align}
  W = X^{k+2} + Y^2 + Z^2\ ,
\end{align}
is well-known to flow in the infrared to the level $k$ $SU(2)/U(1)$  Kazama-Suzuki model,
whose central charge is given by
$c_\text{min}=3\left(1-\frac2k\right)<3$.

The elliptic genus of this minimal model can also be computed by
supersymmetric localization with the result \cite{Witten:1993jg}
\begin{align}
  \CE_\text{min}(\t,z)& = \frac{\vartheta_{1}
  \left(\t,\left(1-\frac1k\right) z\right) }{\vartheta_{1}\left(\t,\frac1k z\right)}\ .
  \label{ellmin}
\end{align}

The $\CN=2$ superconformal character formulae of the
$SU(2)/U(1)$ Kazama-Suzuki model at level $k$ are
\begin{align}
  \text{Ch}^\text{min}_{l,n}(\t,z) =
  \chi_{n,1}^{l}(\t,z) - \chi_{n,3}^l(\t,z)\ ,
\end{align}
where the branching functions $\chi_{m,s}^{l}(\t,z)$ are defined by \cite{Eguchi:2003ik}
\begin{align}
  \chi^{\widehat{\mathfrak{su}}(2)_{k-2}}_l(\t, w) \cdot
  \chi^{\widehat{\mathfrak{u}}(1)_2}_s(\t,w-z)
  = \sum_{n=0}^{2k-1} \chi^{\widehat{\mathfrak{u}}(1)_{k}}_n (\t, w- \frac{2z}{k+2})
  \cdot \chi_{n,s}^{l}(\t,z)\ ,
\end{align}
where $\chi^{\widehat{\mathfrak{su}}(2)_k}_l(\t,z)$ and
$\chi^{\widehat{\mathfrak{u}}(1)_{k}}_n(\t,z)$ denote
the $\widehat{\mathfrak{su}}(2)_k$ and $\widehat{\mathfrak{u}}(1)_k$
characters given by
\begin{align}
  \chi^{\widehat{\mathfrak{su}}(2)_k}_l(\t,z) & =
  \frac{\vartheta_{k+2,l+1}(\t,z/2)- \vartheta_{k+2,-l-1}(\t,z/2)}
  {\vartheta_{2,1}(\t,z/2) - \vartheta_{2,-1}(\t,z/2)}
  \qquad (l=0,1,..,k)\ ,
  \nonumber \\
  \chi^{\widehat{\mathfrak{u}}(1)_{k}}_n(\t,z) & =
  \frac{\vartheta_{k,n}(\t,z/2)}{\eta(q)}
  \qquad (n=0,1,..,2k-1)\ .
\end{align}
The minimal model characters are periodic in $m$ with period $2k$,
\begin{align}
  \text{Ch}^\text{min}_{l,n}(\t,z) = \text{Ch}^\text{min}_{l,n+2k}(\t,z)\ ,
\end{align}
and also enjoy a $\mathbb{Z}_2$ reflection symmetry
\begin{align}
  \text{Ch}^\text{min}_{l,n}(\t,z) = - \text{Ch}^\text{min}_{(k-2)-l,k+n}(\t,z)\ .
\end{align}
Using these two properties, one can always choose $(l,n)$
to satisfy the constraint
\begin{align}
   0 \leq |n \mp 1| \leq l\ .
\end{align}
Then, the conformal weight $h$ and $r$ charge of the highest
weight representation corresponding to $\text{Ch}^\text{min}_{l,n}(\t,z)$
are
\begin{align}
  h_{l,n} - \frac{c_\text{min}}{24} & = \frac{(l+1)^2 - n^2}{4k} \ ,
  \nonumber \\
  r_{l,n}& =- \frac nk \pm \frac12\ .
\end{align}
Under spectral flow by $\a$ units, the $\CN=2$ characters $\text{Ch}^\text{min}_{l,n}$
transform as
\begin{align}
  q^{\frac{c_\text{min}}{6} \a^2 } \left(e^{2\pi i z}\right)^{\frac{c_\text{min}}{3} \a}
  \text{Ch}^\text{min}_{l,n}(\t,z+ \a\t)
  = (-1)^\a \text{Ch}^\text{min}_{l,n+2\a}(\t,z)\ .
\end{align}

We can then express the elliptic genus of the $\CN=2$ $\frac{SU(2)}{U(1)}$
minimal as
\begin{align}
  \CE_\text{min}(\t,z) = \sum_{l=1}^{k-1} \text{Ch}^\text{min}_{l-1,l}(\t,z)\ ,
  \label{minellch}
\end{align}
where the $\CN=2$ characters can be written as
\begin{align}
  \text{Ch}^\text{min}_{l-1,l}(\t,z) =
  \frac{i\vartheta_{1}(\t,z)}{\eta(\t)^3} e^{-2\pi i z \frac{l}{k}}  \sum_{w\in \mathbb{Z}}
  q^{kw^2+ l w } \left[ \frac{1}{1- q^{l+kw} e^{2\pi i z}}+\frac{1}{1- q^{kw} e^{-2\pi i z}} -1 \right]\ .
\end{align}
Note that the $\CN=2$ characters $\text{Ch}^\text{min}_{l-1,l}(\t,z)$
correspond to primary vertex operators
${\tilde V}_{\frac{l-1}{2};\frac{l-1}{2},\frac{l-1}{2}}^\text{susy}\left(\frac12,\frac12\right)$.
This implies that the elliptic genus $\CE_\text{min}$ receives
contributions only from the characters associated with the Ramond ground
states.

\subsection{DSLST}

We saw earlier that the holographic dual of DSLST at the particular point in the moduli space corresponding to the brane configuration of figure \ref{figure1} contains the $\mathbb{Z}_k$ orbifold of the product of an $\CN=2$ cigar SCFT and an $\CN=2$ minimal model:
\begin{align}
  \left( \frac{SL(2)_k}{U(1)} \times \frac{SU(2)_k}{U(1)}
  \right)/{\mathbb{Z}_k}\ .
\end{align}
The $\mathbb{Z}_k$ orbifold action, that is generated by $e^{2\pi i (2J_\text{R}^3)}$ with
\begin{align}
  2J_\text{R}^3 = J_\text{R}^\text{cig} + J_\text{R}^\text{min} \ ,
\end{align}
is necessary for space-time supersymmetry.\footnote{And worldsheet $\CN=4$ superconformal symmetry.}

In the case of this particular class of orbifold
theories, we can use the results of \cite{Kawai:1993jk}
to obtain (see e.g. \cite{Eguchi:2008ct,Ashok:2012qy,Cheng:2014zpa})
\begin{align}
  \CE_\text{DSLST}(\t,z) = \frac1k \sum_{\a,\b=0}^{k-1} q^{\frac{\hat c}{2}\a^2}
  \left( e^{2\pi i z} \right)^{\hat c \a } \CE_\text{cig}(\t,z+\a\t+\b)
  \CE_\text{min}(\t,z+\a\t+\b)\ ,
\end{align}
where the elliptic genera of the two coset models are given by
(\ref{resultccc}) and (\ref{ellmin}), and the central charge is
\begin{align}
  \hat c = \frac{c_\text{cig}}{3} + \frac{c_\text{min}}{3} = 2\ .
\end{align}
Clearly we obtain a non-holomorphic elliptic genus since the cigar
elliptic genus is not holomorphic.
The contribution from the discrete states of DSLST can be read off from (\ref{resultbbb}),
\begin{align}
  \CE_\text{DSLST}^d(\t,z) = \frac1k \sum_{\a,\b=0}^{k-1} q^{\frac{\hat c}{2}\a^2}
  \left( e^{2\pi i z} \right)^{\hat c \a } \CE_\text{cig}^d(\t,z+\a\t+\b)
  \CE_\text{min}(\t,z+\a\t+\b)\ .
\end{align}
Using (\ref{minellch}) and (\ref{cigellch}), it is
also useful to rewrite the discrete part of the elliptic
genus in terms of $\CN=2$ superconformal characters as
\begin{align}
  \CE_\text{DSLST}^d =
  \sum_{\a=0}^{k-1} \sum_{l=1}^{k-1} \sum_{\tilde l=0}^{k-1} \sum_{w\in\mathbb{Z}}
  \delta(l+\tilde l - k)  \cdot
  \text{Ch}^\text{cig}_{\tilde l+1,-\tilde l +2\a+2kw}(\t,z)
  \cdot
  \text{Ch}^\text{min}_{l-1,l+2\a}(\t,z)\ .
  \label{resultN2char}
\end{align}
The $\mathbb{Z}_k$ projection gives rise to the Kronecker delta
in the above expression.

In the next section we discuss various features of the discrete
contribution to the elliptic genus and their physical implications.


\section{Properties of the Elliptic Genus}

\subsection{$\CN=4$ Character Decomposition}

The superconformal field theory appearing in the holographic
description of DSLST has an  $\CN=4$ superconformal algebra with $c=6$.
It must therefore be possible to decompose the discrete contribution to
the elliptic genus into a (in general infinite) sum of $\CN=4$ characters.
The irreducible highest weight representations $V^{(m)}_{h,j}$ of the
$\CN=4$ superconformal algebra with $c=6(m-1)$ are labelled by
$h$ and $j$, the eigenvalues of $L_0$ and  $\left(2J^3_\text{R}\right)_0$.
We define the Ramond sector characters as
\begin{align}
\text{ch}^{(m)}_{h,j}(\t,z)=\text{Tr}_{V^{(m)}_{h,j}}
\left[ (-1)^{F} e^{2\pi i z (2J^3_\text{R})_0} q^{L_0-c/24} \right]\ .
\end{align}
These characters are given by \cite{Eguchi:1987wf}
\begin{align}
  \text{ch}^{(m)}_{\frac{m-1}{4},j}(\t,z) = i \mu^{(m)}_j(\t,z)
  \frac{(\vartheta_1(\tau,z))^2}{\vartheta_1(\t,2z) \cdot \eta(\tau)^3}
\end{align}
for the massless or BPS characters with $h=\frac{m-1}{4}$ and
$j \in \{0,1, \cdots m-1 \}$, and by
\begin{align}
  \text{ch}^{(m)}_{h,j} =i (-1)^{j} q^{h-\frac{m-1}{4}-\frac{j^2}{4m}}
  \Big( \vartheta_{m,j}(\t,z)-\vartheta_{m,-j}(\t,z) \Big)
  \frac{(\vartheta_1(\tau,z))^2}{\vartheta_1(\t,2z) \cdot \eta(\tau)^3}
\end{align}
for the massive or non-BPS characters with $h>\frac{m-1}{4}$ and $j \in \{1,2, \cdots m-1 \}$.
Here the function
\begin{align}
\mu_j^{(m)}(\tau,z)= (-1)^{j+1} \sum_{k \in \IZ} q^{m k^2} \left(e^{2\pi i z}\right)^{2mk} \sum_{a=-j}^{j+1} \frac{\left( e^{2\pi i z} q^k \right)^a}{1- e^{2\pi i z} q^k}\ ,
\end{align}
is a generalized Appell-Lerch sum and for $m=2$ is closely related to the Appell-Lerch sum $\mu(\tau,z)$ that plays
a prominent role in Zwegers influential work on mock theta functions \cite{Zwegers:2008zna}.
One can show that the second Taylor coefficients of $\CN=4$ massless and massive characters are given by
\begin{align}
  \left. \left( \frac{1}{2\pi i}\right)^2 \frac{d^2}{d z^2}
  \text{ch}^{(m)}_{\frac{m-1}{4},0}(\t,z) \right|_{z=0} & =
  4 \sum_{n=1}^\infty \left[ \frac{q^n \left(1- q^{m n^2}\right)}{(1-q^n)^2}
  - mn \frac{ q^{m n^2}\left(1+q^n\right)}{1-q^n} \right]\ ,
  \nonumber \\
  \left. \left( \frac{1}{2\pi i}\right)^2 \frac{d^2}{d z^2}
  \text{ch}^{(m)}_{h,1}(\t,z) \right|_{z=0} & =
  - 2 q^{h-\frac{m-1}{4}} \vartheta_{m,1}^{(1)}(\t)\ .
\end{align}

It is not hard to see that the decomposition into $\CN=4$ characters involves
the massless character $\text{ch}_{\frac14,0}^{(2)}$ with multplicity $(k-1)$. As discussed in Section 4.2, these
correspond under spectral flow to the chiral operators in the NS sector that can be understood as
the relative translation modes of the fivebranes.  In terms of world-volume fields on IIB fivebranes they
belong to the same supermultiplet as the $k-1$ massless gauge bosons in the Cartan subalgebra of $SU(k)$.
Denoting the multiplicities of massive characters by $a_n$ we thus have the decomposition
\begin{align}
  \CE_\text{DSLST}^d(\t,z) = (k-1) \text{ch}_{\frac14,0}^{(2)}(\t,z) + \sum_{n=1}^\infty
  a_n \text{ch}_{\frac14+n,1}^{(2)}(\t,z) \, .
  \label{result10}
\end{align}

Based on non-trivial numerical experimentation, we believe that
the second Taylor coefficient of $\CE_\text{DSLST}^d(\t,z)$ for arbitrary $k$ is
\begin{align}
 \left. \left( \frac{1}{2\pi i}\right)^2 \frac{d^2}{d z^2}
 \CE_\text{DSLST}^d(\t,z) \right|_{z=0} =  4 \CF_2^{k,1}\ ,
\end{align}
and thus the coefficients $a_n$ satisfy the relation
\begin{align}
  - \frac12 \vartheta_{2,1}^{(1)}(\t) \sum_{n=1} a_n q^{ n -\frac18}   =
  \CF_{2}^{k,1}(q) - (k-1) \CF_{2}^{2,1}(q)
\end{align}
with
\begin{align}
  \CF_2^{k,1} & = \left [ \sum_{\substack{r,s\in \mathbb{Z} \\ 0<s<kr}} -
  k \sum_{\substack{r,s\in \mathbb{Z} \\ 0<ks<r}} \right] s q^{rs} =
  \sum_{n=1}^\infty \left[ \frac{q^n \left(1- q^{k n^2}\right)}{(1-q^n)^2}
  - kn \frac{ q^{kn^2}\left(1+q^n\right)}{1-q^n} \right]\ .
\end{align}
Here $\vartheta_{2,1}^{(1)}(\t)$ denotes the first Taylor coefficient
of a Jacobi theta function with level $2$, $\vartheta_{2,1}(\t,z)$,
\begin{align}
  \vartheta_{2,1}^{(1)}(\t) = \sum_{n\in\mathbb{Z}} ( 1 + 4n) q^{\frac{(4n+1)^2}{8}}
  = \eta(q)^3\ .
\end{align}
The $\CF_2^{k,1}$ are mixed mock modular forms of weight two that played
an important role in the analysis of \cite{Harvey:2014cva}.
It is natural to expect a relation between the second derivative of
$\CE_\text{DSLST}^d(\t,z) $ at $z=0$ and the spacetime BPS index computed
in \cite{Harvey:2014cva} since they are both weight two
(mixed) mock modular forms computed in the SCFT describing
the holographic background of DSLST.

For later convenience, we present the first few coefficients $a_n$ ($k>2$) below
\begin{align}
  a_1 = 2k -4 \ \ , \qquad a_2 = 8k - 20 \ , \qquad a_3 = \left\{ \begin{array}{cl} 6 & \text{ if } ~k=3 \\ 22k-66 & \text{ if } ~k> 3 \end{array} \right. \ ,
  \label{result101}
\end{align}
and so on. At $k=2$, all the coefficients $a_n$ vanish and the elliptic genus
is simply given by the $\CN=4$ massless character with $j=0$
\begin{align}
  \CE_\text{DSLST}^d(\t,z) = \text{ch}^{(2)}_{\frac14,0}(\t,z) ~\text { at }~ k=2\ .
  \label{zzzzz}
\end{align}

\subsubsection{Comments on $k=2$}

Note that the $\CN=2$ minimal model contribution to the elliptic genus
is not present at $k=2$. It is therefore natural to ask how our result at $k=2$ is related to the elliptic genus
of the $\mathbb{Z}_2$ orbifold of the cigar theory at $k=2$
studied in \cite{Eguchi:2010cb,Troost:2010ud,Ashok:2011cy}.

Using the results of \cite{Kawai:1993jk},
the elliptic genus of the $\mathbb{Z}_k$ orbifold of the coset CFT takes the general form
\begin{align}
  \CE^{D}_\text{orb}(\t,z) = \frac1k \sum_{\a,\b=0}^{k-1} (-1)^{D(\a+\b+\a \b)}
  e^{2\pi i \frac{\hat c}{2} \a\b} q^{\frac{\hat c}{2} \a^2}
  \left( e^{2\pi i z} \right)^{\hat c \a} \CE_\text{cig}(\t,z+\a\t+\b)\ ,
\end{align}
where $D$ is an integer satisfying
\begin{align}
  D k = \hat c k \text{ mod } 2\ .
  \label{condition1}
\end{align}

For a generic $k$, we can choose $D=1$ satisfying the relation (\ref{condition1}),
\begin{align}
  k = \left(1 + \frac 2k\right) k \text{ mod } 2\ .
\end{align}
and the elliptic genus then becomes
\begin{align}
  \CE^{D=1}_\text{orb}(\t,z) = \frac1k \sum_{\a,\b=0}^{k-1} (-1)^{(\a+\b)}
  e^{2\pi i \frac1k \a\b} q^{\frac{\hat c}{2} \a^2}
  \left( e^{2\pi i z} \right)^{\hat c \a} \CE_\text{cig}(\t,z+\a\t+\b)\ ,
\end{align}
which agrees with the results in \cite{Eguchi:2010cb,Troost:2010ud,Ashok:2011cy}.
When $k=2$, one can show that
\begin{align}
  \CE^{D=1}_\text{orb}(\t,z) & = \frac{i \vartheta_{1}(\t,z)}{\eta(q)^3} \cdot
  \sum_{m\in\mathbb{Z}} \frac{q^{2m^2} \xi^{2m}}{1- \xi^\frac12 q^m}
  \nonumber \\ & =
  \left(1+ \frac{1}{\sqrt{\xi}} \right) + \frac{\left( \sqrt{\xi} + 1 \right)^3
  \left( \sqrt{\xi}- 1 \right)^2}{\xi^{\frac32}} q + \CO(q^2)\ ,
  \label{result1}
\end{align}
where $\xi=e^{2\pi i z}$. However, the above result cannot be decomposed into
$\CN=4$ superconformal characters. This implies that  $\CN=2$ supersymmetry
can not be enhanced to  $\CN=4$ supersymmetry when $k=2$ and $D=1$.
Furthermore, there are states in (\ref{result1}) that carry
fractional $U(1)$ R-charges indicating that the choice $D=1$ leads to a theory
which is not compatible with spacetime supersymmetry of DSLST.

It is interesting to understand where the discrepancy between the
two results (\ref{zzzzz}) and (\ref{result1}) at $k=2$ comes from.
In fact, for $k=2$ we find that there is  another solution to
(\ref{condition1}), namely $D=2$ since
\begin{align}
  D \cdot 2 = \left(1 + \frac{2}{2} \right) \cdot 2 \ .
\end{align}
The corresponding elliptic genus $\CE^{D=2}_\text{orb}(\t,z)$ turns out to coincide
with a single $\CN=4$ massless character with $j=0$, i.e.,
\begin{align}
  \CE^{D=2}_\text{orb}(\t,z) & = \frac12 \sum_{\a,\b=0}^{1}
  q^{\frac{\hat c}{2} \a^2} \left( e^{2\pi i z} \right)^{\hat c \a}
  \CE_\text{cig}(\t,z+\a\t+\b)
  \nonumber \\ & =
  \text{ch}^{(2)}_{\frac14,0}(\t,z)\ ,
\end{align}
which now in turn agrees perfectly with the elliptic genus of double-scaled little
string theory (DSLST) at $k=2$ (\ref{zzzzz}).

\subsubsection{Large $k$ limit}

Consider the discrete contribution to the elliptic genus of DSLST for $k$
fivebranes in the limit $k \rightarrow \infty$.
We might expect that it becomes easier to identify vertex operators
for various states in this limit, which will be discussed in section 4.2, since
the algebraic structure simplifies.

It is not hard to check that
\begin{equation}
\lim_{k \rightarrow \infty} {\cal F}_2^{k,1}(q) = \frac{1-E_2(\t)}{24}
\end{equation}
with $E_2(\t)$ the quasi modular Eisenstein series of weight $2$.
In particular it is independent of $k$ at large $k$. We therefore have, pulling out an
overall factor of $k$,
\begin{equation}
\lim_{k \rightarrow \infty} {\cal E}_\text{DSLST}^d(\tau,z)  =
k  \left( {\rm ch}^{(2)}_{\frac{1}{4},0}(\tau,z)  + \sum_{n=1}^\infty a_n
{\rm ch}^{(2)}_{\frac{1}{4}+n,1}(\tau,z) \right)
\end{equation}
where
\begin{equation}
\sum_{n=1} a_n q^{n-1/8}=   \frac{2}{\eta(\tau)^3} {\cal F}_2^{2,1}(q)
\end{equation}

It is perhaps interesting to rewrite this further using the fact \cite{Dabholkar:2012nd}
that
\begin{equation}
{\cal F}_2^{2,1}(q) = \frac{\eta(\tau)^3 H^{(2)}(\t)}{48} + \frac{E_2(\t)}{24}
\end{equation}
where
\begin{equation}
H^{(2)}(\t)= 2 q^{-1/8} \left( -1 + 45 q + 231 q^2 + 770 q^3 + \cdots \right)
\end{equation}
is the weight $1/2$ mock modular form connected to Mathieu Moonshine \cite{Eguchi:2010ej}
that appears in the decomposition of the the elliptic genus of $K3$ into
$\CN=4$ characters. We thus have
\begin{equation}
\sum_{n=1} a_n q^{n-1/8}=  \frac{H^{(2)}(\t)}{24} + \frac{E_2(\t)}{12 \eta(\tau)^3}
\end{equation}

The elliptic genus of $K3$ has a decomposition into characters of the $\CN=4$ SCA given by
\begin{equation}
\CE_{K3}(\tau,z)= 24 {\rm ch}^{(2)}_{\frac{1}{4},0}(\tau,z) +
\sum_{n=0}^\infty c_n {\rm ch}^{(2)}_{\frac{1}{4}+n,1}(\tau,z)
\end{equation}
with
\begin{equation}
\sum_{n=0}^\infty c_n q^{n-1/8} = H^{(2)}(\tau) \, .
\end{equation}
We thus have
\begin{equation}
\lim_{k \rightarrow \infty} {\cal E}_\text{DSLST}^d(\tau,z)  =
k  \left(  \frac{\CE_{K3}(\tau,z)}{24} + Z_\text{quasi}(\tau,z) \right)
\end{equation}
where
\begin{equation}
Z_\text{quasi}(\tau,z)= \sum_{n=0}^\infty b_n {\rm ch}^{(2)}_{\frac{1}{4}+n,1}(\tau,z)
\end{equation}
and
\begin{equation}
\sum_{n=0}^\infty b_n q^{n-1/8} = \frac{E_2(\t)}{12 \eta(\tau)^3}
\end{equation}

This shows that the large $k$ limit of the DSLST elliptic genus is not modular since $E_2$ is only quasi modular.
It would be interesting to develop a physical interpretation of the above decomposition of the large $k$ limit of the DSLST elliptic genus into a modular part,
proportional to the elliptic genus of $K3$, and a quasi-modular part.


\subsection{Vertex Operators and Null States}

The elliptic genus of DSLST, $\CE_\text{DSLST}$, is independent of the position of the fivebranes (see section 5). Thus, if we make the radius of the circle in figure \ref{figure1}, $R_0$, large, the naive expectation is that the fivebranes do not interact with each other and the elliptic genus should be proportional to $k$.

However we can see from (\ref{result10}) and (\ref{result101}) that the elliptic genus
of DSLST exhibit a more complicated dependence on $k$,
\begin{align}
  \CE_\text{DSLST}^d(\t,z)  & = (k-1) \text{ch}_{\frac14,0}^{(2)}(\t,z) +
  2(k-2) \text{ch}_{\frac14+1,1}^{(2)}(\t,z) + \cdots
  \nonumber \\  & =
  (k-1) \bigg[ 1 + \left( e^{4\pi i z} - 2 e^{2\pi i z} +
  2 - 2 e^{-2\pi i z} + e^{- 4\pi i z} \right) q + \CO(q^2)  \bigg]\ ,
  \nonumber \\  &  +
  2 ( k- 2) \bigg[ \left(-  e^{2\pi i z} +
  2 -  e^{-2\pi i z}  \right) q + \CO(q^2)  \bigg] + \CO(q^2)\ .
  \label{result44}
\end{align}
We will explain in section 5 that this result does not contradict the fact that the elliptic genus
is independent of the positions of the fivebranes. Here we will try to identify the vertex operators
that correspond to the first few terms in (\ref{result44}).

To find vertex operators contributing to the elliptic genus,
the expression (\ref{resultN2char}) in
terms of $\CN=2$ superconformal characters is very useful.
The terms in (\ref{resultN2char}) corresponding to the primary
operators that contribute the $\CN=4$ massless character
$\text{ch}^{(0)}_{\frac14,0}$ are
\begin{align}
  \text{Ch}^\text{cig}_{l+1,l}(\t,z) \text{Ch}^\text{min}_{k-l-1,k+l}(\t,z)
  = 1 + 2 \left( 2 - e^{2\pi i z} - e^{-2\pi i z} \right)q
  + \CO(q^2)
  \label{vertex01}
\end{align}
where $l=2,..,k-2$ and
\begin{align}
  \text{Ch}^\text{cig}_{2,1}(\t,z) \text{Ch}^\text{min}_{k-2,k+1}(\t,z)
  & = 1 + \left( 3 - e^{2\pi i z} - 2e^{-2\pi i z} \right)q
  +\CO(q^2)\ ,
  \nonumber \\
  \text{Ch}^\text{cig}_{k,k-1}(\t,z) \text{Ch}^\text{min}_{0,2k-1}(\t,z)
  & = 1 + \left( 3 - 2e^{2\pi i z} - e^{-2\pi i z} \right)q
  +\CO(q^2)\ .
  \label{vertex02}
\end{align}
From these expressions, we can identify the vertex
operators of the lowest conformal weight $h-\frac{c}{24}=0$ as
\begin{align}
  \CO^{(0)}_{j,0} \equiv V_{j;j+1,j+1}^\text{susy}\big(-\frac12,-\frac12\big)
  \cdot
  \tilde V_{j;j,j}^\text{susy}\big(+\frac12,+\frac12 \big)\ ,\label{rrtt}
\end{align}
where $j=0,\frac12,..,\frac{k-2}{2}$. Note that
the operators $\CO^{(0)}_{j,0}$ are related to
the translational modes of the fivebranes via
spectral flow. This explains why the
$\CN=4$ massless character contributions are
proportional to $(k-1)$ rather than $k$. It is due
to the fact that, as will be discussed in details in
section 5, we need to exclude a non-normalizable
translational mode corresponding to the
center of mass of the system.

The other vertex operators in (\ref{vertex01})
and (\ref{vertex02}) of higher conformal weights
can be obtained by acting with $\CN=2$ superconformal
currents of cigar and minimal CFTs on $\CO^{(0)}_{j,0}$.
For instance, we can show from the OPEs in appendix A that
there are $(2k-3)$ vertex operators of conformal weight
$h-\frac{c}{24} = 1$ and $U(1)_R$ charge $r=+1$
\begin{align}
  \left[ \left(G^+_\text{cig}\right)_{-1} , \CO_{j,0}^{(0)} \right] \propto
  \CO^{(0)}_{j,1} \equiv
  V_{j;j,j+1}^\text{susy}\big(\frac12,-\frac12\big)
  \cdot
  \tilde V_{j;j,j}^\text{susy}\big(+\frac12,+\frac12 \big)\ ,\label{pqpqpq}
\end{align}
where $j=0,\frac12,..,\frac{k-2}{2}$, and
\begin{align}
  \left[ \left(G^+_\text{min}\right)_{-1} , \CO_{j,0}^{(0)} \right] \propto
  {\tilde \CO}^{(0)}_{j,1} \equiv
  V_{j;j+1,j+1}^\text{susy}\big(-\frac12,-\frac12\big)
  \cdot
  \tilde V_{j;j-1,j}^\text{susy}\big(+\frac32,+\frac12 \big)\ ,\label{wwee}
\end{align}
where $j=\frac12,1,..,\frac{k-2}{2}$. Note that superficially the number of states of the form (\ref{pqpqpq}),(\ref{wwee}) should be proportional to $k-1$, like that of the states (\ref{rrtt}). However, this is not the case due to the presence of null states, in this case associated with the action of  $\left(G^+_\text{min}\right)_{-1}$ on $\CO_{j=0,0}^{(0)}$.

The other terms in (\ref{resultN2char}) relevant to
find the vertex operators of conformal weight $h-\frac{c}{24}=1$
and positive $U(1)_R$ charge $r>0$ are
\begin{align}
  \text{Ch}^\text{cig}_{l+1,l+2}(\t,z) \text{Ch}^\text{min}_{k-l-1,k+l+2} =
  \left( e^{4\pi i z} - 2 e^{2\pi i z} + 1 \right) q + \CO(q^2)\ ,
  \label{vertex03}
\end{align}
where $l=1,2,..,k-2$, and
\begin{align}
  \text{Ch}^\text{cig}_{k,k+1}(\t,z) \text{Ch}^\text{min}_{0,1}(\t,z) =
  \left(e^{4\pi iz} - e^{2\pi i z} \right)q + \CO(q^2)\ .
  \label{vertex04}
\end{align}
From these $\CN=2$ characters, it is straightforward to identify
the $(k-1)$ vertex operators of conformal weight $h=1+\frac14$
and $U(1)_R$ charge $r=2$,
\begin{align}
  \CO^{(1)}_{j,2} \equiv V_{j;j+1,j+1}^\text{susy} \big(+\frac12,-\frac12\big) \cdot
  {\tilde V}_{j;j,j}^\text{susy} \big(+\frac32,+\frac12\big)\ ,
\end{align}
where $j=0,\frac12,..,\frac{k-2}{2}$.
These operators are in fact $\CN=4$ descendants
in the massless characters
\begin{align}
  \left[ \left(J_\text{R}^{++}\right)_{-1} , \CO^{(0)}_{j,0}\right]
  = \CO^{(1)}_{j,2}\ ,
\end{align}
which explains why their contributions are proportional to $(k-1)$.
Here $J_\text{R}^{++}=J_\text{R}^1 + i J_\text{R}^2$.
On the other hands, the $(2k-3)$ vertex operators in (\ref{vertex03}) and
(\ref{vertex04}) of conformal weight $h-\frac{c}{24} = 1$ and
R-charge $r=1$ can be obtained from acting with
$G^-_\text{cig}$ and $G^-_\text{min}$ on $\CO^{(2)}_{j,2}$,
\begin{align}
  \left[ \left(G^-_\text{cig}\right)_0 , \CO_{j,2}^{(1)} \right] \propto
  \CO^{(1)}_{j,1} \equiv V_{j;j+2,j+1}^\text{susy} \big(-\frac12,-\frac12\big) \cdot
  {\tilde V}_{j;j,j}^\text{susy} \big(+\frac32,+\frac12\big)\ ,
\end{align}
where $j=0,\frac12,..,\frac{k-2}{2}$, and
\begin{align}
  \left[ \left(G^-_\text{min}\right)_0 , \CO_{j,2}^{(1)} \right] & \propto
  {\tilde \CO}^{(1)}_{j,1} \equiv V_{j;j+1,j+1}^\text{susy} \big(+\frac12,-\frac12\big) \cdot
  {\tilde V}_{j;j+1,j}^\text{susy} \big(+\frac12,+\frac12\big)\ ,
\end{align}
where $j=0,1,..,\frac{k-3}{2}$. In this case there is a null state associated with the action of $\left(G^-_\text{min}\right)_{0}$ on
$\CO_{j=\frac{k-2}{2},0}^{(0)}$.

To summarize, we constructed $(4k-6)$ vertex operators of conformal
weight $h-\frac{c}{24} =1$ and R-charge $r=1$ that contribute to the
elliptic genus. Among them, one can show that certain linear combinations of
$\CO^{(0)}_{j,1}$ and ${\tilde \CO}^{(0)}_{j,1}$, and those of
$\CO^{(1)}_{j,1}$ and ${\tilde \CO}^{(1)}_{j,1}$ for $j=0,\frac12,..,\frac{k-2}{2}$
are in fact $\CN=4$ descendants of the ground state, and thus belong to the
massless character. The remaining $2(k-2)$ linear combinations
of these operators, orthogonal to the above $\CN=4$ descendants,
belong to the massive character $\text{ch}^{(2)}_{1+\frac14,0}(\t,z)$.

\subsection{Density of States at Large Level}

In preparation for a discussion of the black hole/string transition in Section 6 we
now turn to an estimate of the entropy of states contributing to the elliptic genus.
The entropy formula can be read off from the asymptotic behavior of the level
density for highly excited perturbative string BPS states. In other words,
we would like to determine the large level $N$
behavior of $D(N,z)$ defined by
\begin{align}
  D(N,z) = \oint \frac{dq}{2\pi i} \frac{\CE^d_\text{DSLST}(\t,z)}{q^{N+1}}\ ,
  \label{saddle1}
\end{align}
where a small circle around the origin is chosen as a contour.

To evaluate the above contour integral for large $N$,
we first need to know how the discrete part
of the elliptic genus $\CE_\text{DSLST}^d(\t,z)$ behaves
as $q \to 1^-$. It is straightforward to estimate crudely that
$\CE_\text{DSLST}^d$ is asymptotic to
%
\begin{align}
  \CE_\text{DSLST}^d \sim \text{Exp}\left[
  \frac{C(z)}{1-q} \right] \text{ as } q \to 1^-
\end{align}
with
\begin{align}
  C(z) = \Big( \text{Li}_2(1) - \text{Li}_2(\xi)
  + \text{Li}_2(\xi^{1/k}) - \text{Li}_2(\xi^{1-1/k})
  \Big) + \text{ c.c.} \ ,
\end{align}
where $\xi= e^{2\pi iz}$ and $\displaystyle \text{Li}_2(x)= \sum_{m=1}\frac{x^m}{m^2}$.

While we could estimate the asymptotic behavior at any value of $z$,
there is cancellation between fermion and boson states
at $z=0$ while at $z=\frac12$ the elliptic genus is essentially
the partition function in the Ramond-Ramond sector with boson and fermion
states contributing with equal signs. Since physically we are interested
in the density of the total number of states we are most interested
in the asymptotic behavior at $z=\frac12$.

Mathematically we note that by
using the identity
\begin{align}
  \text{Li}_2(e^{2\pi i x}) + \text{Li}_2(e^{-2\pi i x}) =  2\pi^2
  \left( x^2 - x + \frac16 \right)\ ,
\end{align}
we can show that $C(z)$ has a maximum at $z=\frac 12$ and
\begin{align}
  C\Big(z=\frac12\Big) = 2\pi^2 \left( \frac12 - \frac{1}{2k}  \right)\ .
\end{align}

We now continue with the saddle point approximation
to evaluate the above contour integral (\ref{saddle1}) at $z=\frac12$.
One finds the saddle point for $q$ near $1$. Indeed the integrand at $q=1-\e$
\begin{align}
  \text{Exp}\left[ \frac{C(\frac12)}{1-q} - (N+1) \log q \right]
  \simeq \text{Exp}\left[ \frac{C(\frac12)}{\e} + (N+1)\e \right] \ ,
\end{align}
becomes stationary at
\begin{align}
  \e \simeq \sqrt{\frac{C(\frac12)}{N+1}} \text{ as } N \to \infty\ .
\end{align}
Therefore, the leading behavior of the degeneracy at high level $N$ is
\begin{align}
  D\Big(N,\frac12\Big) \simeq e^{2\pi \sqrt{\left( 1-\frac1k\right)N} }\ .
\end{align}

After including the additional contribution from $\mathbb{R}^{1,4} \times S^1$, we
can determine the entropy of Dabholkar-Harvey states at high energy in
DSLST as
\begin{align}
  S_\text{string} = 2\pi \sqrt{\left(1 + 1-\frac1k\right) N}\ .\label{Entropystring}
\end{align}
This agrees with a naive application of the Cardy formula for a theory with $c_\text{eff}=6(2-{1\over k})$, as in \cite{Giveon:2005mi}.


\section{(In)dependence of Moduli}

In our discussion above we focused on the elliptic genus of Little String Theory
at a particular point in its moduli space at which the fivebranes are placed
equidistantly on a circle in the transverse $\mathbb{R}^4$. More general points
in the moduli space correspond to other distributions of fivebranes in $\mathbb{R}^4$ and
it is natural to ask how the answer depends on the positions of the fivebranes.

Superficially, we expect the elliptic genus to be independent of the
moduli since, as we explained before,
it encodes the number of spacetime $1/4$ BPS states with
particular momentum and winding $(P,W)$ on a longitudinal $S^1$.
The mass of these states (\ref{bpsmass}) is independent of the position moduli,
and their degeneracy is an integer that cannot depend on continuous parameters
such as positions of fivebranes.\footnote{A related point is that
one can think of the fivebrane background (\ref{dslstt}) as a non-compact K$3$,
and it is well known that for compact K$3$'s the elliptic genus is independent of the moduli.}

This leaves the possibility of jumps in the number of such states
at some specific values of the moduli, that is  a wall-crossing phenomenon that is known
to occur for some BPS states in field and string theory.
In particular, our analysis above is directly applicable when the string coupling
of DSLST in small, i.e. when the mass of a $D1$-brane stretched between
any two NS$5$-branes is much larger than $m_s$.
As mentioned above, in that regime the states in question are perturbative string states,
quite analogous to the perturbative BPS states studied in \cite{Dabholkar:1989jt}.

One might wonder whether there are possible jumps in the spectrum when the DSLST coupling
is of order one, and the perturbative analysis may receive order one corrections.
We do not expect such jumps when the fivebranes are separated.
In general, the jumps are due to the fact that the
supersymmetric central charges carried by the BPS states depend on the moduli;
they occur when the central charge vectors of different BPS states align.
In our case, the charges carried by the states in question are independent of the
position moduli. As we will see below, the spectrum of BPS states however does exhibit
jumps at points in moduli space where fivebranes collide.
At such points, the spectrum of BPS states goes from that of strings to that of black holes.

\begin{figure}[t]
\centering
\subfigure[]{\includegraphics[width=.4\textwidth]{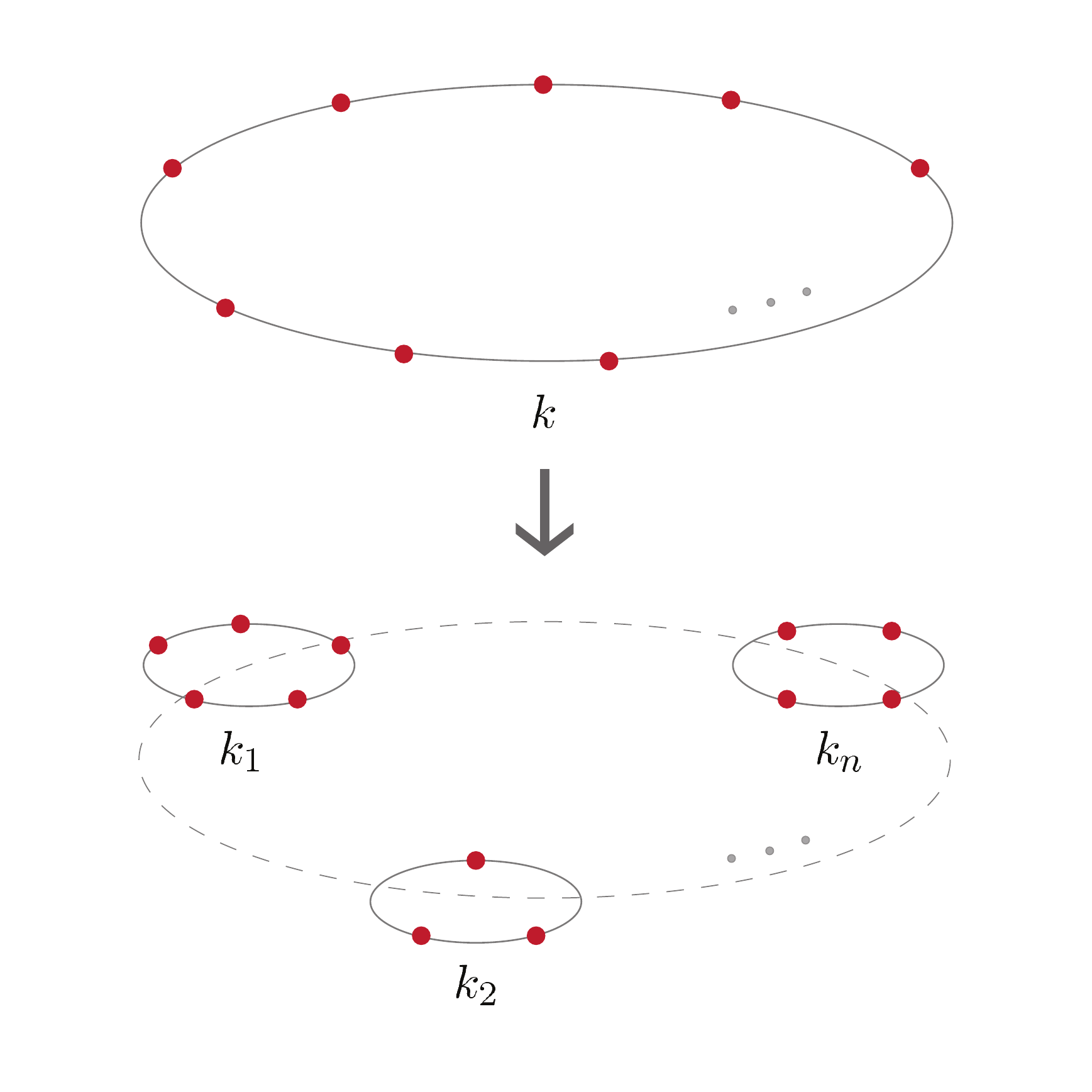}}\qquad
\subfigure[]{\includegraphics[width=.35\textwidth]{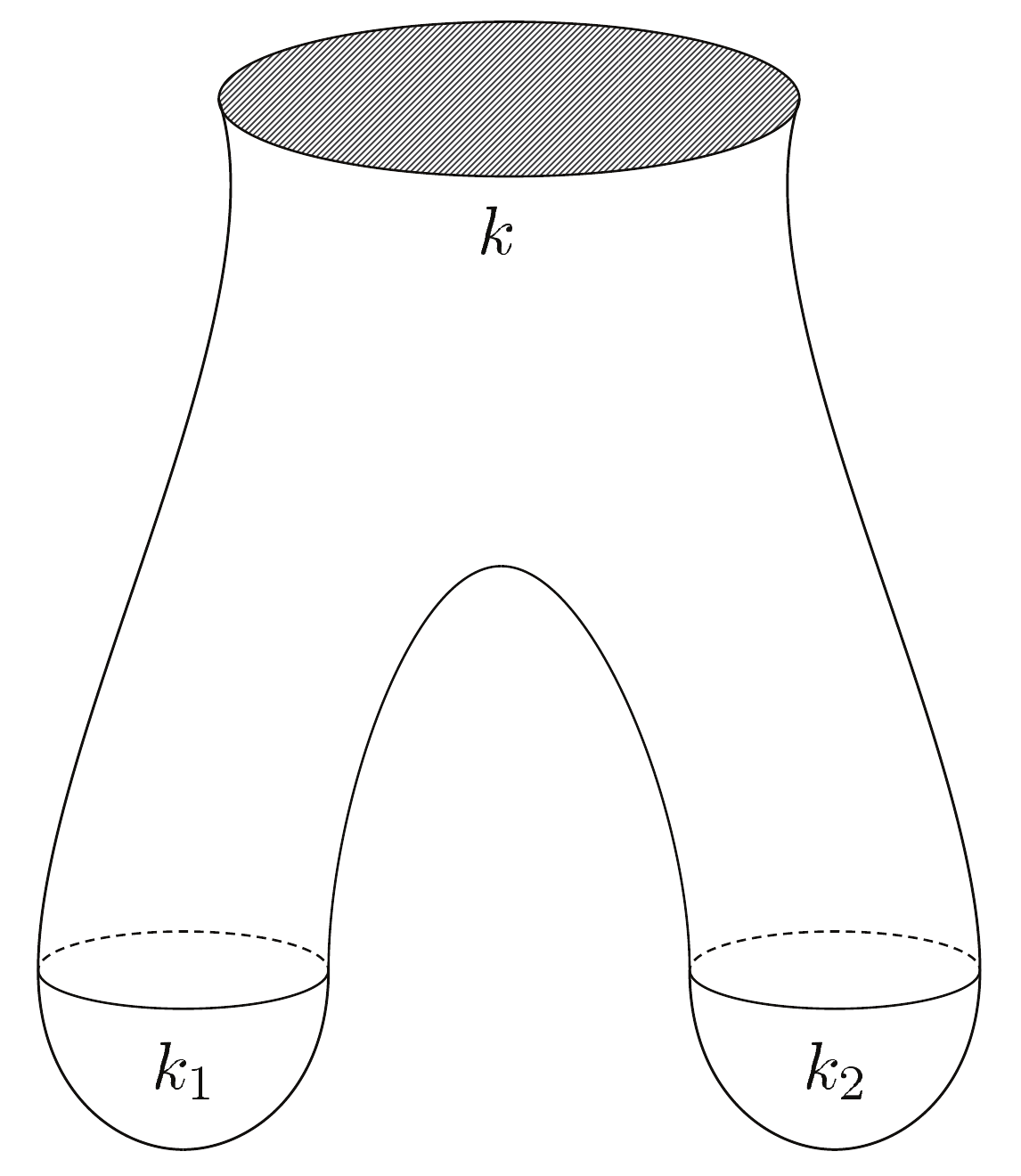}}
\caption{(a) $k$ fivebranes on a single circle are separated
into groups of $(k_1,k_2,..,k_n)$ fivebranes on $n$ circles,
(b) a single throat of $k$ fivebranes is divided into
smaller throats of $k_1$ and $k_2$ fivebranes where $k=k_1+k_2$. }\label{figure2}
\end{figure}
In this section we will stay in the realm of weakly coupled DSLST,
where the spectrum of BPS states is expected to be independent of the positions
of the fivebranes. This independence may seem surprising from the spacetime point of view.
Consider, for example, a deformation that takes the original configuration of
$k$ fivebranes on a circle to one where they are separated into groups of
$(k_1, k_2,\cdots, k_n)$ fivebranes that live on $n$ well separated circles, depicted
in figure \ref{figure2} (a).
The discrete part of the elliptic genus receives contributions
from normalizable states that live in the fivebrane geometry.
Let us denote the number of such states with given $(P,W)$
in the single circle configuration of figure \ref{figure1} by $F_{P,W}(k)$.
If we separate the fivebranes into $n$ circles,
as in figure \ref{figure2} (a), and assume that the states that contribute
to the elliptic genus are localized in the vicinity of the individual circles, then
the number of states in the second configuration should be $\sum_{i=1}^n
F_{P,W}(k_i)$. Since the degeneracy of states with given $(P,W)$ must be independent
of the positions of the fivebranes, we conclude that
\begin{align}
 F_{P,W}(k)=\sum_{i=1}^n F_{P,W}(k_i)  \label{fpwk}
\end{align}
for all $k_i$ satisfying $\sum_{i=1}^n k_i=k$.  However, we saw in previous sections
that the degeneracies computed from the elliptic genus do not actually
satisfy this relation. For instance, see equation (\ref{result44}).
In this section we will discuss the origin of this discrepancy.

While the states we are interested in are $1/4$ BPS, it is useful first to recall
the situation with $1/2$ BPS states. These are the modes that correspond to
the positions of NS$5$-branes in $\mathbb{R}^4$, and their partners under
spacetime supersymmetry.  The translational modes can be viewed as
deformations of the harmonic function
\begin{align}
 H=l_s^2\sum_{j=1}^k {1\over |\vec x-\vec x_j|^2}\ ,  \label{harfn}
\end{align}
which determines the metric, dilaton and NS two-form $B$-field of
fivebranes located at $\vec x=\vec x_j$, $j=1,2,\cdots, k$,
\begin{gather}
 ds^2 = dx_\mu dx^\mu + H(\vec x) d\vec x\cdot d\vec x\ , \nonumber \\
  e^{2(\Phi-\Phi_0)} = H(\vec x)\ , \nonumber \\
 H_{mnp} = -\epsilon^q_{mnp}\partial_q\Phi\ .  \label{chs}
\end{gather}
Thus, they can be thought of as gravitons with wave functions obtained by replacing
$ \vec x_j\to \vec x_j+\vec\delta_j$ in (\ref{harfn}), and expanding in $\vec\delta_j$.

The term in $H(\vec x)$ that goes like $\delta^s$
(or, more precisely, $\delta_{j_1}\delta_{j_2}\cdots\delta_{j_s}$
with the vector indices on $\delta_j$ suppressed), behaves at large $|\vec x|$
like $1/|\vec x|^{s+2}$, i.e. like $1/|\vec x|^s$ relative to the leading $1/|\vec x|^2$ term.
Expanding the gravitational action, taking into account the factor of $\exp(-2\Phi)$
in front of the Einstein term, we see that the behavior of the norm at large $|\vec x|$
is given by
\begin{align}
 \int {d|\vec x|\over|\vec x|} |\vec x|^{2-2s}\ . \label{norm}
\end{align}
Therefore, the $s=1$ perturbation is non-normalizable.
Looking back at (\ref{harfn}), we see that this perturbation
corresponds to displacing the center of mass of the fivebrane system in
$\mathbb{R}^4$. The fact that it is non-normalizable in the near horizon
geometry of the fivebranes was found in the original work \cite{Callan:1991dj},
who showed that the wavefunction of this mode is centered
in the transition region between the near-horizon geometry
and the asymptotically flat space far from the branes.
In terms of LST, this implies that the low energy theory
on $k$ NS$5$-branes in type IIB string theory has gauge group $SU(k)$
rather than $U(k)$.

On the other hand, the modes with $s>1$ are normalizable. Their wavefunctions can be
obtained by performing the expansion described above.
We will not need the details of this expansion, except for the fact that
the corresponding wavefunctions are centered in the region near the fivebranes.
Far from the fivebranes, the wavefunctions decay exponentially in the natural variable
$\ln|\vec x|$.

Using this picture, we can now revisit the question of the
(in)dependence of the spectrum of BPS states on the positions of the fivebranes.
The number of translational modes of the fivebranes and their superpartners
is clearly independent of the positions of the fivebranes.
With the center of mass excluded, it is given by $4(k-1)$.
Superficially, this is inconsistent with the discussion around the equation (\ref{fpwk}),
but now we can resolve the discrepancy.

Consider the multi-circle configuration of fivebranes depicted in
figure \ref{figure2} (a). Following the analysis above, we know that
each cluster of $k_i$ fivebranes gives rise to $4(k_i-1)$ translational modes,
with the center of mass degrees of freedom excluded.
This gives rise to $4\sum_i(k_i-1)=4(k-n)$  modes, that are localized
near the respective circles. The missing $4(n-1)$ modes are easy to
identify -- they correspond to modes that preserve the center of mass
of the whole fivebrane configuration, but not the centers of mass
of the separate groups of $k_i$ fivebranes. As in \cite{Callan:1991dj},
their wave functions are not localized near the individual
circles in figure \ref{figure2} (a).
Such a configuration can be thought of as a single throat of
$k$ fivebranes at large $|\vec x|$ that splits into smaller throats
of the individual groups of fivebranes as $|\vec x|$ decreases, as depicted
in figure \ref{figure2} (b).
The wave functions of the $4(n-1)$ missing multiplets are supported in the
transition regions between the large throat and the smaller ones,
and lead to a violation of the logic that led to (\ref{fpwk}).

So far we discussed the behavior of the $1/2$ BPS states,
which can be identified as Ramond-Ramond ground states leading to
the constant contribution to the elliptic genus
given by (\ref{result44}). In particular, we explained why
their contribution $(k-1)$ is independent of the positions of the fivebranes,
and yet is not proportional to $k$. A key point was that the
wave functions of these states do not satisfy decoupling, i.e.,
if we split the fivebranes into arbitrarily well separated groups,
these groups remain entangled via their centers of mass.

As discussed in previous sections, the full elliptic genus can be decomposed
into the contributions of different representations of the $\CN=4$ superconformal algebra,
which is present everywhere in the moduli space of the multi-fivebrane CFT.
The massless $\CN=4$ character contribution  in (\ref{result44}) corresponds to states
that can be thought of as the $1/2$ BPS states discussed above
acted on by left-moving $\CN=4$ superconformal generators. Thus,
the wave functions of these states are the same as those of the
$1/2$ BPS states, and our discussion of the latter applies directly to them.

Massive $\CN=4$ character contributions are in general more complicated, as can be seen from
(\ref{result44}).  At the special point in the moduli space described by
figure \ref{figure1}, we demonstrated in section 4.2 that these states can be obtained
by acting on the $1/2$ BPS states with left-moving $\CN=2$ generators of the
$SL(2,\mathbb{R})/U(1)$ and $SU(2)/U(1)$ factors. We also found that null
states sometimes can be generated by acting on the Ramond-Ramond ground
states with such $\CN=2$ superconformal generators, which
explains why their contribution is not even proportional to $(k-1)$.

However, at generic points in moduli space,
the chiral algebra of the model is just the $\CN=4$ superconformal algebra,
and such a description is not available.
Nevertheless, we expect the wave functions of these states to have the
same qualitative structure as that of states in the massless $\CN=4$ characters,
for the following reason. All states contributing to the elliptic genus are
Ramond ground states in the right-moving worldsheet sector.
At large $k$, we can think of them as zero modes of the Dirac
equation in the fivebrane background. Squaring this equation gives the
massless Klein-Gordon equation, whose solution is the harmonic function (\ref{harfn}).
Thus, the properties of these states as we change the moduli should be the
same as those in the massless $\CN=4$ representations.


\section{Black Holes versus Perturbative String States}

In the previous section we saw that the contribution of normalizable
LST states to the elliptic genus is independent of the positions of the
fivebranes in $\mathbb{R}^4$. In weakly coupled DSLST these states
can be thought of as
perturbative string states living in the fivebrane background, but
the corresponding spectrum can be extended to regions in moduli space
where the DSLST coupling is of order one.  As mentioned in the previous section,
this picture is expected to be valid for separated fivebranes, but it receives
important modifications when fivebranes are allowed to coincide.

Consider, for example, the configuration of fivebranes on a circle of radius $R_0$
in the transverse space $\mathbb{R}^4$ depicted in figure \ref{figure1}.
For $R_0>0$ we expect the analysis of the previous sections to be valid.
However, for $R_0=0$ there is another competing contribution to the elliptic genus
from a black hole with the same quantum numbers as the perturbative string states
described above.  To construct this black hole, we start with the coincident
fivebrane background \cite{Callan:1991dj},
$\mathbb{R}^{4,1}\times S^1\times \mathbb{R}_\phi\times SU(2)_k$, and look for
solutions that carry the two charges $P$ (momentum) and $W$ (winding)
along the $S^1$ of radius $R$ that the fivebranes wrap.
In string theory, it is convenient to label these charges in terms of
left and right moving momenta,
\begin{align}
  (P_L,P_R)=\left({P\over R}-{WR\over\alpha'}\, ,
  {P\over R}+{WR\over\alpha'}\right)\ .  \label{charges}
\end{align}
For $P_L=P_R=0$, the black hole solution takes the form
$SL(2,\mathbb{R})_k/U(1)\times \mathbb{R}^4\times S^1\times SU(2)_k$.
It describes the background of $k$ non-extremal fivebranes,
with the value of the dilaton at the horizon labeling the energy density
above extremality. For general $(P_L,P_R)$, one can find the
black hole solution by reduction of the three-dimensional rotating,
charged black string background obtained from the uncharged black hole
solution by a sequence of boosts and T dualities.
Algebraically, this leads to a  CFT in which the
$SL(2,\mathbb{R})_k/U(1)\times S^1$ factor above is replaced by
${SL(2,\mathbb{R})_k\times U(1)\over U(1)}$, where
the embedding of the gauged $U(1)$ into $U(1)\times U(1)\subset SL(2,\mathbb{R})\times U(1)$
is determined by the charges $(P_L,P_R)$.
For a review of this construction, as well as for the precise sigma-model
background fields in the general two-charge case,
see e.g. \cite{Giveon:2005mi,Giveon:2006pr}.

A tractable special case, which has all the essential ingredients is  $P_L=0$.
The corresponding charged black hole has metric, dilaton and gauge field,
\begin{align}
ds^2&=-fdt^2+{k\alpha'\over4}{dr^2\over r^2f}\ ,
\nonumber \\
\Phi & =-{1\over 2}\ln\left(\sqrt{k\over \alpha'}\, r\right)\ ,
\nonumber \\
A_t & =\frac{\a'}{2r} P_R \ ,
\label{BHlinD}
\end{align}
where the function $f(r)$ is
\begin{align}
f=\left(1-{r_-\over r}\right)\left(1-{r_+\over r}\right),
\end{align}
and the inner and outer horizons of the black hole are at
\begin{align}
r_\pm= \frac{\a'}{2} \left( M_\text{BH} \pm \sqrt{M_\text{BH}^2-P_R^2} \right)\ .
\end{align}
The entropy of this black hole, and its generalization to generic $(P_L,P_R,M_\text{BH})$
is given by \cite{Giveon:2005mi}
\begin{align}
  S_\text{BH}=\pi l_s\sqrt{k}
  \left(\sqrt{M_\text{BH}^2-P_L^2}+\sqrt{M_\text{BH}^2-P_R^2}\right).\label{bhentr}
\end{align}
For the extremal case $M_\text{BH}=|P_R|$, this takes the form
\begin{align}
  S_\text{BH}=2\pi\sqrt{k PW}\ ,
  \label{BHentropy}
\end{align}
familiar from studies of three-charge black holes.
Looking back at the analogous expression for perturbative strings
(\ref{Entropystring}) of the same charges $(P_L,P_R)$,
\begin{align}
  S_\text{string}=2\pi\sqrt{ \Big( 2-\frac1k \Big) PW}\ ,
  \label{Entropystring2}
\end{align}
we see that the two are qualitatively similar,
but the factor $(2-{1\over k})$ in the fundamental string entropy
is replaced by $k$ for black holes.
Thus, the black hole entropy is always larger for $k\ge 2$.

We conclude that as the fivebranes approach each other,
the number of $1/4$ BPS LST states jumps from
(\ref{Entropystring2}) to (\ref{BHentropy}).
At first sight this is rather surprising -- the positions of the fivebranes
are moduli in the theory, and can be thought of as Higgsing the
$SU(k)$ gauge group to $U(1)^{k-1}$. When the fivebranes are nearly coincident,
the symmetry breaking scale, namely the mass of W-bosons corresponding to
D-strings stretched between NS$5$-branes, becomes very low.
We would not expect it to influence the physics of very massive states,
such as the BPS states contributing to the elliptic genus.
Indeed, in a local QFT such a phenomenon could not occur.
However, LST is not a local QFT, and the states we are interested in
can probe the non-locality; e.g., T-duality, which is often cited
as evidence for non-locality of LST, acts non-trivially on them.
We therefore believe that the jump in the spectrum of $1/4$ BPS
states is an example of UV/IR mixing in LST. The Higgs scale (IR) influences
the spectrum of very massive BPS states (UV).

Another element of the above discussion that we need to address concerns the (non) compactness of the worldvolume of
the fivebranes. Above, we took it to be $\mathbb{R}^{4,1}\times S^1$, but this leads to the following issue.  We see from (\ref{BHlinD}) that the two dimensional string coupling is determined by the mass of the (extremal) black hole,
\begin{align}
e^{-2\Phi(r_\pm)} = \sqrt{\frac{k}{2}} M_\text{BH}\ ,
\end{align}
where we set $\a'=2$ for simplicity. If the four dimensional space along the fivebranes, $\CM_4$, is non-compact, the six dimensional string coupling in the directions along the fivebranes is infinite. Hence, the coset description cannot be studied at small string coupling.

To avoid these singularities it is convenient to compactify
the worldvolume $\mathbb{R}^4$ to $\mathbb{T}^4$. However,
this raises another issue that needs to be addressed.
When the fivebrane worldvolume is taken to be $\mathbb{T}^4\times S^1$,
the LST in question lives in $1+0$ dimensions.
Thus, the moduli corresponding to positions of fivebranes
cannot be taken as fixed, but are rather fluctuating quantum mechanical
degrees of freedom. The states of the theory are characterized
by wave functions on the classical moduli space.
This leads to the question what is the correct interpretation
of our results above in these low dimensional vacua of LST.

Our view on this is that compactified LST has a
discrete set of vacua labeled by the number of coincident fivebranes,
which ranges from $0$ to $k$
(or, more generally, the numbers of coincident fivebranes $(k_1,\cdots, k_n)$
with $\sum_i k_i=k$). The vacuum with no coincident fivebranes has
an elliptic genus that was computed in previous sections.
It can be thought of as due to perturbative string states in the
separated fivebrane background.
The elliptic genus of the vacuum with $k$ coincident fivebranes,
defined formally as the object counting spacetime $1/4$ BPS states,
is dominated by the contribution of the black hole described in this section.
For intermediate numbers of coincident fivebranes we have a combination of the two effects.

Note that the above picture is reminiscent of, but not identical to,
the one discussed in \cite{Witten:1997yu}.
There, the fivebranes were always coincident and the strings
were part of the background. The issue was what is the $1+1$ dimensional
low energy theory on the system of strings and fivebranes, and it was argued
that it splits into Coulomb and Higgs branch CFT's with different central charges.
The Coulomb branch corresponds to strings propagating in the vicinity of, but outside the fivebranes. The Higgs branch describes the theory of strings dissolved in the
fivebranes as self-dual Yang-Mills instantons.
An important role in their separation is played by the fivebrane throat of \cite{Callan:1991dj}
seen by strings propagating in the fivebrane background.

In our case, the only branes in the background are the fivebranes.
We are interested in the full theory rather than just its low energy limit,
and the different branches of the theory correspond to different numbers of coincident
fivebranes. However, the difference in the entropy of BPS states between (\ref{Entropystring2}) and (\ref{BHentropy})
is  the same as in \cite{Witten:1997yu}.
In the Higgs branch this can be read off the Cardy formula with central charge $c=6k$.
In the Coulomb branch one has $c=6$ but the object that governs the
high energy density of states is $c_\text{eff}=6(2-\frac1k)$. The fivebrane throat plays an important role in our discussion as well since in a sector of the Hilbert space of LST with $W\not=0$, we effectively have strings propagating in the vicinity of the fivebranes, as in \cite{Witten:1997yu}.

Another closely related phenomenon is the string-black hole transition
discussed in \cite{Giveon:2005mi}.
There, it was shown that the high energy spectrum of string theory
in asymptotically linear dilaton vacua (i.e. vacua of LST) is dominated for $k>1$
by black holes, while for $k<1$ the black holes are non-normalizable,
and the spectrum is that of perturbative strings.
The dependence of the entropy on the slope of the linear dilaton, $Q$,
that can be parametrized by the number of fivebranes $k$ in this paper
via the relation $Q=\sqrt{\alpha'/k}$ for strings and black holes
is precisely the same as in our analysis above.
However, unlike in \cite{Giveon:2005mi}, we work at a fixed $k>1$, and the
transition between strings and black holes in our case
is between different branches of the theory of $k$ fivebranes.
It would be interesting to understand the relation between the two phenomena better.


\section{Non-Extremal Case}

In the previous sections we saw that LST on $\mathbb{T}^4\times S^1$ exhibits
a non-trivial vacuum structure. Classically,
the number of $1/4$ BPS states carrying the charges (\ref{charges})
jumps when fivebranes coincide. Quantum mechanically,
the theory splits into distinct sectors labeled by the numbers of coincident fivebranes,
each with its own spectrum of BPS states.

In this section we would like to briefly comment on the physics of near-BPS states
in this theory. Consider a point in the moduli space of LST at which the fivebranes
are separated, such that the string coupling is everywhere small.
In that case, we can compute the entropy of near-BPS states using perturbative
string techniques. It is given by
\begin{align}
S_{\rm string}=\pi  l_s\sqrt{2-{1\over k}}
 \left(\sqrt{M_\text{BH}^2-P_L^2}+\sqrt{M_\text{BH}^2-P_R^2}\right).\label{fundstr}
\end{align}
In the BPS case, $M_\text{BH}=|P_R|$, this reduces to (\ref{Entropystring2}).

The energy above the BPS bound effectively makes the fivebranes non-extremal.
Thus, their gravitational attraction exceeds the repulsion due to their $B_{\mu\nu}$ charge,
and the moduli associated with their positions develop an attractive potential.
Thus, the problem becomes time-dependent. However, if the system is near-BPS,
the timescale associated with the motion of the fivebranes towards each other is large,
and one can treat the problem in the adiabatic approximation.

 In particular, there is a long time period in which the thermodynamics of the system
 is that of fundamental strings, and the entropy is given by (\ref{fundstr}).
 Eventually, as the fivebranes get closer, the effective string coupling
 becomes of order one. In this regime, the time evolution becomes rapid
 and the adiabatic approximation breaks down.

When $t\to\infty$, the fivebranes approach each other, and
the effective string coupling in their vicinity becomes large.
In this limit, one can again do thermodynamics, but this time it is governed by
the black hole solutions discussed in the previous section.
The corresponding entropy is given by (\ref{bhentr}),
which has the same form as (\ref{fundstr}), with $2-\frac 1k \to k$.
As expected, the entropy increases with time.

The discussion of the non-extremal case above is useful for understanding
the jump in the BPS entropy discussed in the previous sections.
In the non-extremal problem the behavior of the entropy is a smooth function of
the two parameters in the problem: time, $t$, and the energy above extremality,
$\epsilon$. We start with the system of separated fivebranes $(M_W\not=0)$,
and take the limit $\epsilon\to 0$ (the BPS limit) and $t\to\infty$.
This limit can be taken in two ways that give different answers.
If we first take $\epsilon\to 0$, and then $t\to\infty$,
we get the fundamental string entropy  (\ref{fundstr}).
On the other hand, if we take $t\to\infty$ first and then $\epsilon\to 0$,
we get the black hole answer (\ref{bhentr}).
Slightly away from extremity, the time dependence interpolates smoothly between the two.


\section{Discussion}

In this paper we saw that string theory in a background with $k$
NS5-branes wrapping $\mathbb{T}^4\times S^1$ has non-trivial vacuum structure.
We studied the spectrum of $1/4$ BPS states in the different vacua
and saw that when the fivebranes are coincident,
they can be described as black holes carrying the relevant charges,
while along the Coulomb branch they correspond to pertubrative string
states in the separated fivebrane geometry.

We computed the degeneracies of the two kinds of states, and found that
they do not agree. In particular, at the origin we found the entropy  (\ref{BHentropy}),
while along the Coulomb branch it was (\ref{Entropystring2}).
We interpreted this discrepancy as due to a non-trivial vacuum
structure of the fivebrane theory.

We pointed out that this phenomenon is counterintuitive,
since the origin is a finite distance away from points
along the Coulomb branch, and the metric on
the Coulomb branch does not receive quantum corrections.
In terms of the theory of the fivebranes, known as Little String Theory,
it is possibly due to the non-locality of the theory
and it implies that it exhibits UV/IR mixing.\footnote{Other manifestations of UV/IR mixing in LST have been studied in \cite{Giveon:2015cma,Ben-Israel:2015mda}.}

Our analysis is based on the elliptic genus of DSLST at the special point in the moduli
space that allows the weakly-coupled solvable CFT description (\ref{temp1}). Using
various properties of the elliptic genus discussed in section 4, we argued
in section 5 that the elliptic genus is independent of the positions of the NS5-branes.
This result is consistent with the fact that, when the LST is defined on
$\mathbb{T}^4\times S^1$, the notion of classical moduli space is not
well defined and the ground state of the theory
can be characterized by a wavefunction on the position moduli space.
Finally, we obtained the degeneracy of such ground states (\ref{Entropystring}) from
the asymptotic behavior of the elliptic genus at large level.

Our results have implications to other issues.
One is the program to describe microstates of supersymmetric
black holes in terms of horizonless geometries \cite{Mathur:2005zp,Bena:2013dka}.
The quarter BPS black holes that figured in our analysis
are nothing but the three-charge black holes whose microstates
are discussed in that program. Usually, these black holes are studied in the full,
asymptotically flat, spacetime of string theory.
However, as discussed in the present work,
one can also study them in the near-horizon geometry of the
fivebranes, which is an asymptotically linear dilaton spacetime.

The main idea of the microstate program is to find geometries
that look asymptotically far from the horizon like the corresponding black hole,
but that deviate from it near the location of the would-be horizon,
and in particular do not have a horizon themselves.
The hope is that the entropy of these horizonless geometries
agrees with the Bekenstein entropy of the black hole.

Our results point to a subtlety with this program.
We saw that when the fivebranes are separated,
even by a small distance, the BPS states can be thought of as standard
fundamental string states in the smooth background of the fivebranes.
One can describe these states by vertex operators in the fivebrane background,
but one can also write the supergravity fields around the strings that
carry momentum and winding.  In flat spacetime this was done in
\cite{Callan:1995hn,Dabholkar:1995nc}, and a similar construction
should work in the fivebrane background.

The supergravity fields around these fundamental strings are
presumably essentially the same as those describing the black hole
solution with the same charges, at least at large distance
from the horizon. Thus, one might be tempted to think of them as
microstates of the black hole. However, the picture we were led to
in this paper is different. The horizonless geometries corresponding
to the fundamental string states in the separated fivebrane background
and the black hole are different objects. In fact, they live in
different vacua  of the fivebrane theory, and their entropies are not the same.
Thus, our results suggest that a horizonless geometry that
approximates well the black hole geometry outside of the would be horizon
can not necessarily be thought of as a microstate of the black hole.

Our point of view is compatible with that of \cite{Sen:2009bm},
where it was argued that horizonless geometries and black holes with the
same quantum numbers correspond to different states.
In that case the different descriptions were valid in different duality frames,
i.e. different regions in coupling space,
whereas for us the black holes and fundamental strings describe the
BPS states in different vacua of the same theory.
Our picture also seems to be related to that of \cite{Martinec:2014gka},
although the precise relation remains to be understood.

Our discussion in section 7 is reminiscent of other phenomena in string theory.
For example, the authors of \cite{Horowitz:1996nw} discussed the
transition from fundamental strings to black holes that happens
as a function of the string coupling. In particular,
they argued that if one considers a typical highly excited
fundamental string state, and continuously raises the string coupling,
at some point the  Schwarzschild radius of a black hole
with the same mass and charges as the fundamental
string exceeds the string scale, and the fundamental string description
gives way to a black hole one.

Something similar happens dynamically in our system.
If we start with non-extremal fivebranes in the region where
the effective LST string coupling is small,
the entropy is dominated by fundamental string states.
As time goes by, the fivebranes approach each other,
the effective string coupling grows, and at late time
the system is better described as a black hole. Thus,
our system can be used to study the transition of \cite{Horowitz:1996nw}
in a controlled setting.

Another related problem is that discussed in \cite{Atick:1988si}.
These authors studied the thermodynamics of weakly coupled string theory
in asymptotically flat spacetime, and emphasized that due to the Jeans instability,
thermodynamics does not really make sense in this system.
Rather, at any finite density the system will develop time dependence.
However, if the time variation is sufficiently slow,
one can still study weakly coupled string thermodynamics,
and the resulting description is valid for a long time.

The bulk description of our system (in terms of an asymptotically linear dilaton spacetime)
is similar to that of \cite{Atick:1988si}.
Away from extremality, the system is time dependent,
but if the fivebranes are sufficiently well separated and
the non-extremality is sufficiently small,
the time evolution is slow. Thus the fundamental string picture
is valid for a long time, but it eventually breaks down when the fivebranes
get close and the system makes a transition to a black hole phase.
In our case we have an alternative description of the dynamics in terms
of fivebrane physics (due to LST holography),
and one can use it to shed additional light on the discussion of \cite{Atick:1988si}.

The discussion of this paper can be generalized in a number of directions.
We studied the vacuum structure of six dimensional LST,
but one could generalize the analysis to other dimensional vacua of LST, e.g.
those studied in \cite{Giveon:1999zm}.
There are reasons to believe that the study of such vacua involves
new qualitative and quantitative phenomena.

In our analysis, the degeneracy of BPS states on the Coulomb
branch was obtained by studying string propagation in the
fivebrane background. It is natural to ask whether the
results could alternatively be obtained from the holographically dual point of view.
(IIB) LST reduces in the IR to six dimensional $\CN=(1,1)$ supersymmetric
Yang-Mills theory, and it would be interesting to
see how much of the structure we found can be understood in that theory,
e.g. along the lines of \cite{Kim:2015gha}.

\section*{Acknowledgments}

We thank Nissan Itzhaki, Kimyeong Lee, Emil Martinec and Piljin Yi for discussions.
The work of AG was supported in part by the BSF -- American-Israel Bi-National Science Foundation, by the
I-CORE Program of the Planning and Budgeting Committee and the Israel Science Foundation (Center No.
1937/12), and by a center of excellence supported by the Israel Science Foundation (grant number 1989/14).
JH acknowledges the support of NSF grant 1214409. DK and SL are supported in part by DOE grant DE-FG02-13ER41958. DK thanks Tel Aviv University and  the Hebrew University for hospitality during some stages of this work.  AG and DK thank the organizers of the 8$^\text{th}$ Crete Regional Meeting in String Theory  for the opportunity to present these and related results. SL thanks KIAS and Virginia Tech for hospitality, where part of this work was done.

\newpage

\centerline{\Large \bf Appendix}

\appendix


\section{Review on Coset CFTs}\label{cosetappendix}

\subsection{Cigar CFT}

The supersymmetric $SL(2)_k$ WZW model
can be described by $SL(2)$ currents $J^i$ and three free fermions
$\psi^i$ ($i=1,2,3$) satisfying the OPEs below
\begin{align}
  J^i(z) J^j(0) & \sim \frac{\frac k2 \eta^{ij}}{z^2}+ i \e^{ijk} \frac{J_k(0)}{z}
  \nonumber \\
  J^i(z) \psi^j(0) & \sim i \e^{ijk} \frac{\psi_k(0)}{z}
  \nonumber \\
  \psi^i(z) \psi^j(0) & \sim \frac{\eta^{ij}}{z}\ ,
\end{align}
where $\eta^{ij}=\text{diag}(+1,+1,-1)$.
Let us define a new $SL(2,\mathbb{R})$ current $j^i$,
\begin{align}
  j^i = J^i + \frac{i}{2}\e^{ijk} \psi_j \psi_k\ .
  \label{defcurrent1}
\end{align}
One can then show that the currents $j^i$ commute with three fermions
$\psi^i$ and generate a bosonic $SL(2)$ WZW model at level $k+2$.
Let us define $\psi^\pm = \frac{1}{\sqrt2}\left(\psi^1+i\psi^2\right)$
for later convenience.

The supersymmetric $SL(2)/U(1)$ coset model can be
obtained by gauging the $U(1)$ $\CN=1$ supermultiplet that contains
the primary $\l^3$ and $J^3=\{ G_{-1/2}, \l_3 \}$. Then the coset
has an {\it enhanced} $\CN=2$ algebra generated by
\begin{align}
  G^\pm_\text{cig} & = \sqrt{\frac{2}{k}} j^\mp \psi^{\pm}
  \nonumber \\
  J_R^\text{sl} & = \frac{k+2}{k} \psi^+ \psi^- + \frac 2k j^3 = \l^+\l^- + \frac 2k J^3\ .
  \label{defcurrent2}
\end{align}
%

We denote by $x$, $H$, $X$, and $X_R$ the bosonizations of various currents
$j^3$, $\psi^+\psi^-$, $J^3$ and $J_R$,
\begin{align}
  j^3 &  = - \sqrt{\frac{k+2}{2}} \partial x \ ,
  \nonumber \\
  \psi^+\psi^- & = + i \partial H\ ,
  \nonumber \\
  J^3 & = - \sqrt{\frac{k}{2}} \partial X \ ,
  \nonumber \\
  J_R & = + i \sqrt{ \frac{k+2}{k}} \partial X_R\ .
  \label{boscurrentcigar}
\end{align}
Note that two $U(1)$ currents $J^3$ and $J^R$ commute.
From (\ref{defcurrent1}) and (\ref{defcurrent2}), one can show that
\begin{align}
  x & = \sqrt{ \frac{k+2}{k} } X + i \sqrt{\frac 2k} X_R \ ,
  \nonumber \\
  i H & = \sqrt{\frac 2k} X + i \sqrt{\frac{k+2}{k}} X_R \ .
\end{align}
Using the non-compact parafermion fields $\pi,\pi^\dagger$, the ladder operators $j^\pm$ then can be expressed as follows
\begin{align}
  j^+ & = \sqrt{k+2} \cdot \pi(z) \cdot e^{-\sqrt{\frac{2}{k+2}} x(z)} \ ,
  \nonumber \\
  j^- & = \sqrt{k+2} \cdot \pi^\dagger(z) \cdot  e^{+\sqrt{\frac{2}{k+2}} x(z)} \ .
\end{align}
The two supercurrent $G^\pm_\text{cig}$ can be then expressed as
\begin{align}
  G^+_\text{cig} & = \sqrt{\frac{2(k+2)}{k}} \cdot \pi^\dagger(z)
  \cdot e^{i\sqrt{\frac{k}{k+2}}  X_R(z)}\ ,
  \nonumber\\
  G^-_\text{cig} & = \sqrt{\frac{2(k+2)}{k}} \cdot  \pi(z)
  \cdot e^{-i\sqrt{\frac{k}{k+2}}  X_R(z)}\ ,
\end{align}

\paragraph{Vertex Operators}
Let us discuss the primaries of the coset model.
We start with the $SL(2)$ vertex operator $\Phi^\text{sl}_{j;m,\bar m}$
of conformal weight $-\frac{j(j+1)}{k}$. One can define the $SL(2)/U(1)$
vertex operator $V_{j;m,\bar m}^\text{susy}(\a,\bar \a)$
by removing the $U(1)_{J^3}$ part
of the operator $\Phi^\text{sl}_{j;m,\bar m}$,
\begin{align}
  e^{i\a H} e^{i \bar \a \bar H } \Phi_{j;m,\bar m}^\text{sl} \equiv
  e^{\sqrt{\frac 2k} \left( (m+\a) X + (\bar m+\bar \a) \bar X \right) }
  V^\text{susy}_{j;m,\bar m}(\a,\bar \a) \ .
\end{align}
The conformal weight and $U(1)$ R-charge of $V_{j;m,\bar m}^\text{susy}(\a,\bar \a)$ are
\begin{align}
  h & = \frac{(m+\a)^2-j(j+1)}{k}  + \frac12 \a^2 \ ,
  \nonumber \\
  \bar h & = \frac{(\bar m+\bar \a)^2 - j(j+1)}{k} + \frac12 \bar{\a}^2 \ ,
\end{align}
and
\begin{align}
  r & = \frac{2(m+\a)}{k} + \a\ ,
  \nonumber \\
  \bar r & = \frac{2(\bar m+ \bar \a)}{k} + \bar \a \ .
\end{align}

From the definition of the primary operator $V^\text{B}_{j;m,\bar m}$ for
the bosonic $SL(2)/U(1)$ at level $k+2$
\begin{align}
  \Phi^\text{sl}_{j;m,\bar m}
  \equiv
  e^{\sqrt{\frac{2}{k+2}} (m x  + \bar m \bar x ) }
  V^\text{B}_{j;m,\bar m} \ ,
\end{align}
we have
\begin{align}
  V_{j;m,\bar{m}}^\text{susy}(\a,\bar \a) =
  V_{j;m,\bar{m}}^\text{B} e^{i \frac{2}{\sqrt{k(k+2)}} \left(
  m X_R + \bar{m} \bar{X}_R\right)}
  e^{i \sqrt{\frac{k+2}{k}} \left( \a  X_R + \bar\a \bar{X}_R\right)}\ ,
\end{align}
where $X_R$ denotes the bosonization of the $U(1)$ R-current $J_R^\text{sl}$
(\ref{boscurrentcigar}).
From the well-known equivalence between the non-compact parafermionic
primaries,
\begin{align}
  V^\text{B}_{j;m,\bar m} =
  V^\text{B}_{\frac{k-2}{2}-j;\frac{k+2}{2}+m,\frac{k+2}{2}+\bar m}\ ,
\end{align}
we can verify an interesting property that $V^\text{susy}_{j;m,\bar m}(\a,\bar \a)$
should satisfy
\begin{align}
  V^\text{susy}_{j;m,\bar m}(\a,\bar \a) = V^\text{susy}_{\frac{k-2}{2}-j;\pm \frac{k+2}{2}+m,
  \pm \frac{k+2}{2}+\bar m}(\a\mp1,\bar\a\mp 1)\ .
\end{align}

\paragraph{Useful OPEs}  Finally, let us summarize several useful OPEs
for primaries. The parafermionic primaries $V_{j;m,\bar m}^\text{B}$
satisfy the following OPEs
\begin{align}
  \pi(z) V^B_{j;m,\bar m}(0) &  \sim \frac{m+(j+1)}{\sqrt{k+2}} \frac{1}{z^{1-\frac{2m}{k+2}}}
  V_{j;m+1,\bar m}(0)\ ,
  \nonumber \\
  \pi^\dagger(z) V^B_{j;m,\bar m}(0) & \sim
  \frac{m-(j+1)}{\sqrt{k+2}} \frac{1}{z^{1+\frac{2m}{k+2}}}
  V_{j;m-1,\bar m}(0)\ .
\end{align}
Then, one can easily show that
\begin{align}
  G^+_\text{cig}(z) V^\text{susy}_{j;m,\bar m}(\a,\bar\a)(0)
  & \sim \frac{m-(j+1)}{z^{1-\a}}
  \sqrt{\frac 2k} V^\text{susy}_{j;m-1,\bar m}(\a+1,\bar\a)
  +\cdots \ ,
  \nonumber \\
  G^-_\text{cig}(z) V^\text{susy}_{j;m,\bar m}(\a,\bar\a)(0)
  & \sim \frac{m+(j+1)}{z^{1+\a}}
  \sqrt{\frac 2k} V^\text{susy}_{j;m+1,\bar m}(\a-1,\bar\a)
  +\cdots \ .
\end{align}
%

\subsection{Minimal Model}

The supersymmetric $SU(2)_k$ WZW model
can be described by $SU(2)$ currents $\tilde J^a$ and three free fermions
$\psi^a$ ($a=1,2,3$) satisfying the OPEs below
\begin{align}
  \tilde J^a(z) \tilde J^b(0) & \sim \frac{\frac k2 \delta^{ab}}{z^2}+ i \e^{abc} \frac{\tilde J_c(0)}{z}
  \nonumber \\
  \tilde J^a(z) \psi^b(0) & \sim i \e^{abc} \frac{\psi_c(0)}{z}
  \nonumber \\
  \psi^a(z) \psi^b(0) & \sim \frac{\delta^{ab}}{z}\ ,
\end{align}
where $\delta^{ab}=\text{diag}(+1,+1,+1)$.
Let us define a new $SU(2)$ current $\tilde j^a$,
\begin{align}
  \tilde j^a = \tilde J^a + \frac{i}{2}\e^{abc} \psi_b \psi_c\ .
  \label{defcurrent3}
\end{align}
One can then show that the currents $\tilde j^a$ commute with three fermions
$\psi^a$ and generate a bosonic $SU(2)$ WZW model at level $k-2$.

The supersymmetric $SU(2)/U(1)$ coset model can be
obtained by gauging the $U(1)$ $\CN=1$ supermultiplet that contains
the primary $\psi^3$ and $\tilde J^3=\{ G_{-1/2}, \psi_3 \}$. Then the coset
has an {\it enhanced} $\CN=2$ algebra generated by
\begin{align}
  G^\pm_\text{min} & = \sqrt{\frac{2}{k}} \tilde j^\mp \psi^{\pm}
  \nonumber \\
  J_R^\text{su} & = \frac{k-2}{k} \psi^+ \psi^- - \frac 2k \tilde j^3
  = \psi^+\psi^- - \frac 2k \tilde J^3\ .
  \label{defcurrent4}
\end{align}
%

For later convenience, let us denote by $\tilde x$, $\tilde H$,
$\tilde X$, and $\tilde X_R$ the bosonizations of various currents
$\tilde j^3$, $\psi^+\psi^-$, $\tilde J^3$
and $J_R^\text{su}$,
\begin{align}
  \tilde j^3 &  = i \sqrt{\frac{k-2}{2}} \partial \tilde x \ ,
  \nonumber \\
  \psi^+\psi^- & =  i \partial \tilde H\ ,
  \nonumber \\
  \tilde J^3 & = i \sqrt{\frac{k}{2}} \partial \tilde X \ ,
  \nonumber \\
  J_R^\text{su} & = + i \sqrt{ \frac{k-2}{k}} \partial \tilde X_R\ .
  \label{boscurrentmin}
\end{align}
Note that two $U(1)$ currents $\tilde J^3$ and $J_R^\text{su}$ commute.
From (\ref{defcurrent3}) and (\ref{defcurrent4}), one can show that
\begin{align}
  \tilde x & = \sqrt{ \frac{k-2}{k} } \tilde X -  \sqrt{\frac 2k} \tilde X_R \ ,
  \nonumber \\
  \tilde H & = \sqrt{\frac 2k} \tilde X +  \sqrt{\frac{k-2}{k}} \tilde X_R \ .
\end{align}
Using the  parafermion fields $\tilde \pi,\tilde \pi^\dagger$,
the ladder operators $\tilde j^\pm$ then can be expressed as follows
\begin{align}
  \tilde j^+(z) & = \sqrt{k-2} \cdot \tilde \pi(z) \cdot e^{+i\sqrt{\frac{2}{k-2}} \tilde x}(z) \ ,
  \nonumber \\
  \tilde j^-(z) & = \sqrt{k-2} \cdot \tilde \pi^\dagger(z) \cdot  e^{-i\sqrt{\frac{2}{k+2}} \tilde x}(z) \ .
\end{align}
The two supercurrent $G^\pm_\text{min}$ can be also written as
\begin{align}
  G^+_\text{min} & = \sqrt{\frac{2(k-2)}{k}} \cdot \tilde \pi^\dagger(z)
  \cdot e^{+i\sqrt{\frac{k}{k-2}}  \tilde X_R(z)}\ ,
  \nonumber\\
  G^-_\text{min} & = \sqrt{\frac{2(k-2)}{k}} \cdot  \tilde \pi(z)
  \cdot e^{-i\sqrt{\frac{k}{k-2}} \tilde X_R(z)}\ ,
\end{align}

\paragraph{Vertex Operator} Let us then discuss the primaries of the
supersymmetric $SU(2)/U(1)$ coset model. We start with the
$SU(2)_{k-2}$ vertex operator $\Phi^\text{su}_{\tilde j;\tilde m,\bar{\tilde m}}$
of conformal weight $\frac{\tilde j(\tilde j+1)}{k}$.
One can obtain the $SU(2)/U(1)$
vertex operator ${\tilde V}_{\tilde j;\tilde m,\bar{\tilde m}}^\text{susy}(\b,\bar \b)$
by removing the $U(1)_{\tilde J^3}$ part
of the operator $\Phi^\text{su}_{\tilde j;\tilde m,\bar{\tilde m}}$,
\begin{align}
  e^{i\b \tilde H} e^{i \bar \b \bar{\tilde H} } \Phi_{\tilde j;\tilde m,\bar{\tilde m}}^\text{su} \equiv
  e^{i \sqrt{\frac 2k} \left( (\tilde m+\b) \tilde X + (\bar{\tilde m}+\bar \b) \bar{\tilde X} \right) }
  {\tilde V}^\text{susy}_{\tilde j;\tilde m,\bar{\tilde m}}(\b,\bar \b) \ .
\end{align}
The conformal weight and $U(1)$ R-charge of
${\tilde V}^\text{susy}_{\tilde j;\tilde m,\bar{\tilde m}}(\b,\bar \b) $ are
\begin{align}
  h & = \frac{\tilde j(\tilde j+1) - (\tilde m+\b)^2 }{k}  + \frac12 \b^2 \ ,
  \nonumber \\
  \bar h & = \frac{ \tilde j(\tilde j+1) - (\bar{\tilde m}+\bar \b)^2 }{k} + \frac12 \bar{\b}^2 \ ,
\end{align}
and
\begin{align}
  r & = \frac{2(\tilde m+\b)}{k} + \b\ ,
  \nonumber \\
  \bar r & = \frac{2(\bar{\tilde m}+ \bar \b)}{k} + \bar \b \ .
\end{align}

From the definition of the primary operator ${\tilde V}^\text{B}_{\tilde j;\tilde m,\bar{\tilde m}}$ for
the bosonic $SU(2)/U(1)$ at level $k-2$
\begin{align}
  \Phi^\text{su}_{\tilde j;\tilde m,\bar{\tilde m}}
  \equiv
  e^{i \sqrt{\frac{2}{k-2}} (\tilde m \tilde x  + \bar{\tilde m} \bar{\tilde x} ) }
  {\tilde V}^\text{B}_{\tilde j;\tilde m,\bar{\tilde m}} \ ,
\end{align}
we have
\begin{align}
  {\tilde V}_{\tilde j;\tilde m,\bar{\tilde m}}^\text{susy}(\b,\bar \b) =
  {\tilde V}_{\tilde j;\tilde m,\bar{\tilde m}}^\text{B} e^{-i \frac{2}{\sqrt{k(k-2)}} \left(
  \tilde m \tilde X_R + \bar{\tilde m} \bar{\tilde X}_R\right)}
  e^{i \sqrt{\frac{k-2}{k}} \left( \b \tilde X_R + \bar\b \bar{\tilde X}_R\right)}
\end{align}
where $\tilde X_R$ denotes the bosonization of the $U(1)$ R-current $J_R^\text{su}$ (\ref{boscurrentmin}).
From the well-known equivalence between the non-compact parafermionic
primaries,
\begin{align}
  {\tilde V}^\text{B}_{\tilde j;\tilde m,\bar{\tilde m}} =
  {\tilde V}^\text{B}_{\frac{k-2}{2}-\tilde j;-\frac{k-2}{2}+\tilde m,-\frac{k-2}{2}+\bar{\tilde m}}\ ,
\end{align}
we can verify an interesting property that $\Psi^\text{B}_{\tilde j;\tilde m,\bar{\tilde m}}$
should satisfy
\begin{align}
  {\tilde V}^\text{susy}_{\tilde j;\tilde m,\bar{\tilde m}}(\b,\bar \b) =
  {\tilde V}^\text{susy}_{\frac{k-2}{2}-\tilde j; \pm \frac{k-2}{2} +\tilde m,
  \pm \frac{k-2}{2} +\bar{\tilde m}}
  (\b\pm1,\bar \b\pm1)\ .
\end{align}

\paragraph{Useful OPEs}  Finally, let us summarize several useful OPEs
for primaries. The parafermionic primaries ${\tilde V}_{j;m,\bar m}^\text{B}$
satisfy the following OPEs
\begin{align}
  \tilde \pi(z) {\tilde V}^B_{\tilde j;\tilde m,\bar{\tilde m}}(0) &
  \sim \frac{\tilde j-\tilde m}{\sqrt{k-2}} \frac{1}{z^{1+\frac{2\tilde m}{k-2}}}
  {\tilde V}^B_{\tilde j;\tilde m+1,\bar{\tilde m}}(0)\ ,
  \nonumber \\
  \tilde \pi^\dagger(z) {\tilde V}^B_{\tilde j;\tilde m,\bar{\tilde m}}(0) & \sim
  \frac{\tilde j + \tilde m}{\sqrt{k-2}} \frac{1}{z^{1-\frac{2\tilde m}{k-2}}}
  {\tilde V}^B_{\tilde j;\tilde m-1,\bar{\tilde m}}(0)\ .
\end{align}
Then, one can easily show that
\begin{align}
  G^+_\text{min} (z) {\tilde V}^\text{susy}_{\tilde j;\tilde m,\bar{\tilde m}}(\b,\bar\b) (0)
  & \sim \frac{\tilde j + \tilde m}{z^{1-\b}} \sqrt{\frac 2k} {\tilde V}^\text{susy}_{\tilde j;\tilde m-1,\bar{\tilde m}}(\b+1,\bar\b) (0)
  +\cdots \ ,
  \nonumber \\
  G^-_\text{min} (z) {\tilde V}^\text{susy}_{\tilde j;\tilde m,\bar{\tilde m}}(\b,\bar\b) (0)
  & \sim \frac{\tilde j - \tilde m}{z^{1+\b}} \sqrt{\frac 2k} {\tilde V}^\text{susy}_{\tilde j;\tilde m+1,\bar{\tilde m}}(\b-1,\bar\b) (0)
  +\cdots \ .
\end{align}

\newpage

\bibliographystyle{JHEP}
\bibliography{Refs}

\def\cprime{$'$}
\providecommand{\href}[2]{#2}\begingroup\raggedright\begin{thebibliography}{10}

\bibitem{Strominger:1996sh}
A.~Strominger and C.~Vafa, {\it {Microscopic origin of the Bekenstein-Hawking
  entropy}},  {\em Phys. Lett.} {\bf B379} (1996) 99--104,
  [\href{http://arxiv.org/abs/hep-th/9601029}{{\tt hep-th/9601029}}].

\bibitem{Mandal:2010cj}
I.~Mandal and A.~Sen, {\it {Black Hole Microstate Counting and its Macroscopic
  Counterpart}},  {\em Nucl. Phys. Proc. Suppl.} {\bf 216} (2011) 147--168,
  [\href{http://arxiv.org/abs/1008.3801}{{\tt arXiv:1008.3801}}]. [Class.
  Quant. Grav.27,214003(2010)].

\bibitem{Mathur:2005zp}
S.~D. Mathur, {\it {The Fuzzball proposal for black holes: An Elementary
  review}},  {\em Fortsch. Phys.} {\bf 53} (2005) 793--827,
  [\href{http://arxiv.org/abs/hep-th/0502050}{{\tt hep-th/0502050}}].

\bibitem{Bena:2013dka}
I.~Bena and N.~P. Warner, {\it {Resolving the Structure of Black Holes:
  Philosophizing with a Hammer}},  \href{http://arxiv.org/abs/1311.4538}{{\tt
  arXiv:1311.4538}}.

\bibitem{Giveon:2005mi}
A.~Giveon, D.~Kutasov, E.~Rabinovici, and A.~Sever, {\it {Phases of quantum
  gravity in AdS(3) and linear dilaton backgrounds}},  {\em Nucl. Phys.} {\bf
  B719} (2005) 3--34, [\href{http://arxiv.org/abs/hep-th/0503121}{{\tt
  hep-th/0503121}}].

\bibitem{Giveon:2005jv}
A.~Giveon and D.~Kutasov, {\it {The Charged black hole/string transition}},
  {\em JHEP} {\bf 01} (2006) 120,
  [\href{http://arxiv.org/abs/hep-th/0510211}{{\tt hep-th/0510211}}].

\bibitem{Dabholkar:1989jt}
A.~Dabholkar and J.~A. Harvey, {\it {Nonrenormalization of the Superstring
  Tension}},  {\em Phys. Rev. Lett.} {\bf 63} (1989) 478.

\bibitem{Berkooz:1997cq}
M.~Berkooz, M.~Rozali, and N.~Seiberg, {\it {Matrix description of M theory on
  T$^4$ and T$^5$}},  {\em Phys. Lett.} {\bf B408} (1997) 105--110,
  [\href{http://arxiv.org/abs/hep-th/9704089}{{\tt hep-th/9704089}}].

\bibitem{Seiberg:1997zk}
N.~Seiberg, {\it {New theories in six-dimensions and matrix description of M
  theory on T$^5$ and T$^5/Z(2)$}},  {\em Phys. Lett.} {\bf B408} (1997)
  98--104, [\href{http://arxiv.org/abs/hep-th/9705221}{{\tt hep-th/9705221}}].

\bibitem{Losev:1997hx}
A.~Losev, G.~W. Moore, and S.~L. Shatashvili, {\it {M \& m's}},  {\em Nucl.
  Phys.} {\bf B522} (1998) 105--124,
  [\href{http://arxiv.org/abs/hep-th/9707250}{{\tt hep-th/9707250}}].

\bibitem{Aharony:1999ks}
O.~Aharony, {\it {A Brief review of 'little string theories'}},  {\em Class.
  Quant. Grav.} {\bf 17} (2000) 929--938,
  [\href{http://arxiv.org/abs/hep-th/9911147}{{\tt hep-th/9911147}}].

\bibitem{Kutasov:2001uf}
D.~Kutasov, {\it {Introduction to little string theory}},  in {\em
  {Superstrings and related matters. Proceedings, Spring School, Trieste,
  Italy, April 2-10, 2001}}, pp.~165--209, 2001.

\bibitem{Aharony:1998ub}
O.~Aharony, M.~Berkooz, D.~Kutasov, and N.~Seiberg, {\it {Linear dilatons, NS
  five-branes and holography}},  {\em JHEP} {\bf 10} (1998) 004,
  [\href{http://arxiv.org/abs/hep-th/9808149}{{\tt hep-th/9808149}}].

\bibitem{Callan:1991dj}
C.~G. Callan, Jr., J.~A. Harvey, and A.~Strominger, {\it {World sheet approach
  to heterotic instantons and solitons}},  {\em Nucl. Phys.} {\bf B359} (1991)
  611--634.

\bibitem{Callan:1991at}
C.~G. Callan, Jr., J.~A. Harvey, and A.~Strominger, {\it {Supersymmetric string
  solitons}},  \href{http://arxiv.org/abs/hep-th/9112030}{{\tt
  hep-th/9112030}}.

\bibitem{Giveon:1999px}
A.~Giveon and D.~Kutasov, {\it {Little string theory in a double scaling
  limit}},  {\em JHEP} {\bf 10} (1999) 034,
  [\href{http://arxiv.org/abs/hep-th/9909110}{{\tt hep-th/9909110}}].

\bibitem{Giveon:1999tq}
A.~Giveon and D.~Kutasov, {\it {Comments on double scaled little string
  theory}},  {\em JHEP} {\bf 01} (2000) 023,
  [\href{http://arxiv.org/abs/hep-th/9911039}{{\tt hep-th/9911039}}].

\bibitem{Troost:2010ud}
J.~Troost, {\it {The non-compact elliptic genus: mock or modular}},  {\em JHEP}
  {\bf 1006} (2010) 104, [\href{http://arxiv.org/abs/1004.3649}{{\tt
  arXiv:1004.3649}}].

\bibitem{Eguchi:2010cb}
T.~Eguchi and Y.~Sugawara, {\it {Non-holomorphic Modular Forms and
  $SL(2,R)/U(1)$ Superconformal Field Theory}},  {\em JHEP} {\bf 1103} (2011)
  107, [\href{http://arxiv.org/abs/1012.5721}{{\tt arXiv:1012.5721}}].

\bibitem{Ashok:2011cy}
S.~K. Ashok and J.~Troost, {\it {A Twisted Non-compact Elliptic Genus}},  {\em
  JHEP} {\bf 1103} (2011) 067, [\href{http://arxiv.org/abs/1101.1059}{{\tt
  arXiv:1101.1059}}].

\bibitem{Harvey:2014nha}
J.~A. Harvey, S.~Lee, and S.~Murthy, {\it {Elliptic genera of ALE and ALF
  manifolds from gauged linear sigma models}},  {\em JHEP} {\bf 02} (2015) 110,
  [\href{http://arxiv.org/abs/1406.6342}{{\tt arXiv:1406.6342}}].

\bibitem{Murthy:2013mya}
S.~Murthy, {\it {A holomorphic anomaly in the elliptic genus}},  {\em JHEP}
  {\bf 06} (2014) 165, [\href{http://arxiv.org/abs/1311.0918}{{\tt
  arXiv:1311.0918}}].

\bibitem{Ashok:2013pya}
S.~K. Ashok, N.~Doroud, and J.~Troost, {\it {Localization and real Jacobi
  forms}},  {\em JHEP} {\bf 04} (2014) 119,
  [\href{http://arxiv.org/abs/1311.1110}{{\tt arXiv:1311.1110}}].

\bibitem{Giveon:2014hfa}
A.~Giveon, N.~Itzhaki, and J.~Troost, {\it {Lessons on Black Holes from the
  Elliptic Genus}},  {\em JHEP} {\bf 04} (2014) 160,
  [\href{http://arxiv.org/abs/1401.3104}{{\tt arXiv:1401.3104}}].

\bibitem{Akhoury:1984pt}
R.~Akhoury and A.~Comtet, {\it {Anomalous Behavior of the Witten Index: Exactly
  Soluble Models}},  {\em Nucl. Phys.} {\bf B246} (1984) 253.

\bibitem{Ashok:2014nua}
S.~K. Ashok, E.~Dell'Aquila, and J.~Troost, {\it {Higher Poles and Crossing
  Phenomena from Twisted Genera}},  {\em JHEP} {\bf 08} (2014) 087,
  [\href{http://arxiv.org/abs/1404.7396}{{\tt arXiv:1404.7396}}].

\bibitem{Pioline:2015wza}
B.~Pioline, {\it {Wall-crossing made smooth}},  {\em JHEP} {\bf 04} (2015) 092,
  [\href{http://arxiv.org/abs/1501.01643}{{\tt arXiv:1501.01643}}].

\bibitem{Giveon:2003wn}
A.~Giveon, A.~Konechny, A.~Pakman, and A.~Sever, {\it {Type 0 strings in a 2-d
  black hole}},  {\em JHEP} {\bf 10} (2003) 025,
  [\href{http://arxiv.org/abs/hep-th/0309056}{{\tt hep-th/0309056}}].

\bibitem{Giveon:2015cma}
A.~Giveon, N.~Itzhaki, and D.~Kutasov, {\it {Stringy Horizons}},  {\em JHEP}
  {\bf 06} (2015) 064, [\href{http://arxiv.org/abs/1502.03633}{{\tt
  arXiv:1502.03633}}].

\bibitem{Ben-Israel:2015mda}
R.~Ben-Israel, A.~Giveon, N.~Itzhaki, and L.~Liram, {\it {Stringy Horizons and
  UV/IR Mixing}},  \href{http://arxiv.org/abs/1506.07323}{{\tt
  arXiv:1506.07323}}.

\bibitem{EZ}
M.~Eichler and D.~Zagier, {\em {The Theory of Jacobi Forms}}.
\newblock {Birkh\"auser}, {1985}.
\newblock {Progress in Mathematics (Book 55)}.

\bibitem{Eguchi:2004yi}
T.~Eguchi and Y.~Sugawara, {\it {$SL(2,R) / U(1)$ supercoset and elliptic
  genera of noncompact Calabi-Yau manifolds}},  {\em JHEP} {\bf 0405} (2004)
  014, [\href{http://arxiv.org/abs/hep-th/0403193}{{\tt hep-th/0403193}}].

\bibitem{Israel:2004jt}
D.~Israel, A.~Pakman, and J.~Troost, {\it {D-branes in N=2 Liouville theory and
  its mirror}},  {\em Nucl. Phys.} {\bf B710} (2005) 529--576,
  [\href{http://arxiv.org/abs/hep-th/0405259}{{\tt hep-th/0405259}}].

\bibitem{Witten:1993jg}
E.~Witten, {\it {On the Landau-Ginzburg description of N=2 minimal models}},
  {\em Int. J. Mod. Phys.} {\bf A9} (1994) 4783--4800,
  [\href{http://arxiv.org/abs/hep-th/9304026}{{\tt hep-th/9304026}}].

\bibitem{Eguchi:2003ik}
T.~Eguchi and Y.~Sugawara, {\it {Modular bootstrap for boundary $N = 2$
  Liouville theory}},  {\em JHEP} {\bf 01} (2004) 025,
  [\href{http://arxiv.org/abs/hep-th/0311141}{{\tt hep-th/0311141}}].

\bibitem{Kawai:1993jk}
T.~Kawai, Y.~Yamada, and S.-K. Yang, {\it {Elliptic genera and N=2
  superconformal field theory}},  {\em Nucl. Phys.} {\bf B414} (1994) 191--212,
  [\href{http://arxiv.org/abs/hep-th/9306096}{{\tt hep-th/9306096}}].

\bibitem{Eguchi:2008ct}
T.~Eguchi, Y.~Sugawara, and A.~Taormina, {\it {Modular Forms and Elliptic
  Genera for ALE Spaces}},  in {\em {Workshop on Exploration of New Structures
  and Natural Constructions in Mathematical Physics: On the Occasion of
  Professor Akhiro Tsuchiya's Retirement Nagoya, Japan, March 5-8, 2007}},
  2008.
\newblock \href{http://arxiv.org/abs/0803.0377}{{\tt arXiv:0803.0377}}.

\bibitem{Ashok:2012qy}
S.~K. Ashok and J.~Troost, {\it {Elliptic Genera of Non-compact Gepner Models
  and Mirror Symmetry}},  {\em JHEP} {\bf 07} (2012) 005,
  [\href{http://arxiv.org/abs/1204.3802}{{\tt arXiv:1204.3802}}].

\bibitem{Cheng:2014zpa}
M.~C.~N. Cheng and S.~Harrison, {\it {Umbral Moonshine and K3 Surfaces}},  {\em
  Commun. Math. Phys.} {\bf 339} (2015), no.~1 221--261,
  [\href{http://arxiv.org/abs/1406.0619}{{\tt arXiv:1406.0619}}].

\bibitem{Eguchi:1987wf}
T.~Eguchi and A.~Taormina, {\it {Character Formulas for the $N=4$
  Superconformal Algebra}},  {\em Phys. Lett.} {\bf B200} (1988) 315.

\bibitem{Zwegers:2008zna}
S.~Zwegers, {\em {Mock Theta Functions}}.
\newblock PhD thesis, 2008.
\newblock \href{http://arxiv.org/abs/0807.4834}{{\tt arXiv:0807.4834}}.

\bibitem{Harvey:2014cva}
J.~A. Harvey, S.~Murthy, and C.~Nazaroglu, {\it {ADE Double Scaled Little
  String Theories, Mock Modular Forms and Umbral Moonshine}},  {\em JHEP} {\bf
  05} (2015) 126, [\href{http://arxiv.org/abs/1410.6174}{{\tt
  arXiv:1410.6174}}].

\bibitem{Dabholkar:2012nd}
A.~Dabholkar, S.~Murthy, and D.~Zagier, {\it {Quantum Black Holes, Wall
  Crossing, and Mock Modular Forms}},
  \href{http://arxiv.org/abs/1208.4074}{{\tt arXiv:1208.4074}}.

\bibitem{Eguchi:2010ej}
T.~Eguchi, H.~Ooguri, and Y.~Tachikawa, {\it {Notes on the K3 Surface and the
  Mathieu group $M\_{24}$}},  {\em Exper. Math.} {\bf 20} (2011) 91--96,
  [\href{http://arxiv.org/abs/1004.0956}{{\tt arXiv:1004.0956}}].

\bibitem{Giveon:2006pr}
A.~Giveon and D.~Kutasov, {\it {Fundamental strings and black holes}},  {\em
  JHEP} {\bf 01} (2007) 071, [\href{http://arxiv.org/abs/hep-th/0611062}{{\tt
  hep-th/0611062}}].

\bibitem{Witten:1997yu}
E.~Witten, {\it {On the conformal field theory of the Higgs branch}},  {\em
  JHEP} {\bf 9707} (1997) 003, [\href{http://arxiv.org/abs/hep-th/9707093}{{\tt
  hep-th/9707093}}].

\bibitem{Callan:1995hn}
C.~G. Callan, J.~M. Maldacena, and A.~W. Peet, {\it {Extremal black holes as
  fundamental strings}},  {\em Nucl. Phys.} {\bf B475} (1996) 645--678,
  [\href{http://arxiv.org/abs/hep-th/9510134}{{\tt hep-th/9510134}}].

\bibitem{Dabholkar:1995nc}
A.~Dabholkar, J.~P. Gauntlett, J.~A. Harvey, and D.~Waldram, {\it {Strings as
  solitons and black holes as strings}},  {\em Nucl. Phys.} {\bf B474} (1996)
  85--121, [\href{http://arxiv.org/abs/hep-th/9511053}{{\tt hep-th/9511053}}].

\bibitem{Sen:2009bm}
A.~Sen, {\it {Two Charge System Revisited: Small Black Holes or Horizonless
  Solutions?}},  {\em JHEP} {\bf 05} (2010) 097,
  [\href{http://arxiv.org/abs/0908.3402}{{\tt arXiv:0908.3402}}].

\bibitem{Martinec:2014gka}
E.~J. Martinec, {\it {The Cheshire Cap}},  {\em JHEP} {\bf 03} (2015) 112,
  [\href{http://arxiv.org/abs/1409.6017}{{\tt arXiv:1409.6017}}].

\bibitem{Horowitz:1996nw}
G.~T. Horowitz and J.~Polchinski, {\it {A Correspondence principle for black
  holes and strings}},  {\em Phys. Rev.} {\bf D55} (1997) 6189--6197,
  [\href{http://arxiv.org/abs/hep-th/9612146}{{\tt hep-th/9612146}}].

\bibitem{Atick:1988si}
J.~J. Atick and E.~Witten, {\it {The Hagedorn Transition and the Number of
  Degrees of Freedom of String Theory}},  {\em Nucl. Phys.} {\bf B310} (1988)
  291--334.

\bibitem{Giveon:1999zm}
A.~Giveon, D.~Kutasov, and O.~Pelc, {\it {Holography for noncritical
  superstrings}},  {\em JHEP} {\bf 10} (1999) 035,
  [\href{http://arxiv.org/abs/hep-th/9907178}{{\tt hep-th/9907178}}].

\bibitem{Kim:2015gha}
J.~Kim, S.~Kim, and K.~Lee, {\it {Little strings and T-duality}},
  \href{http://arxiv.org/abs/1503.07277}{{\tt arXiv:1503.07277}}.

\end{thebibliography}\endgroup

\end{document}